\documentclass[a4paper,twocolumn,english,aps,prb,amsmath,showpacs,amssymb]{revtex4}
\pdfoutput=1
\usepackage{ae,aecompl}
\usepackage[T1]{fontenc}
\usepackage[latin9]{inputenc}
\usepackage{amsmath}
\usepackage{amssymb}
\usepackage{graphicx}
\usepackage{esint}

\makeatletter

\pdfpageheight\paperheight
\pdfpagewidth\paperwidth

\providecommand{\tabularnewline}{\\}

\@ifundefined{textcolor}{}
{%
 \definecolor{BLACK}{gray}{0}
 \definecolor{WHITE}{gray}{1}
 \definecolor{RED}{rgb}{1,0,0}
 \definecolor{GREEN}{rgb}{0,1,0}
 \definecolor{BLUE}{rgb}{0,0,1}
 \definecolor{CYAN}{cmyk}{1,0,0,0}
 \definecolor{MAGENTA}{cmyk}{0,1,0,0}
 \definecolor{YELLOW}{cmyk}{0,0,1,0}
 }

\@ifundefined{definecolor}
 {\usepackage{color}}{}
\usepackage[pdftex]{hyperref}
\@ifundefined{definecolor}
 {\usepackage{color}}{}
\usepackage{amsbsy}\usepackage{bm}\usepackage{wasysym}\usepackage{marvosym}

\makeatother

\usepackage{babel}
\begin{document}

\title{Zigzag graphene nanoribbon edge reconstruction with Stone-Wales defects}

\author{J. N. B. Rodrigues$^{1}$, P. A. D. Gonçalves$^{2}$, N. F. G. Rodrigues$^{2}$,
R. M. Ribeiro$^{2}$, J. M. B. Lopes dos Santos$^{1}$ and N. M. R.
Peres$^{2,3}$}

\affiliation{$^{1}$ CFP and Departamento de Física, Faculdade de Ciências Universidade
do Porto, P-4169-007 Porto, Portugal}

\affiliation{$^{2}$Departamento de Física and Centro de Física, Universidade
do Minho, P-4710-057, Braga, Portugal}

\affiliation{$^{3}$Graphene Research Centre and
            Department of Physics, National University of Singapore,
            2 Science Drive 3, Singapore 117542}


\date{\today}
\begin{abstract}
In this article, we study zigzag graphene nanoribbons with edges reconstructed
with Stone-Wales defects, by means of an empirical (first-neighbor)
tight-binding method, with parameters determined by \emph{ab-initio
}calculations of very narrow ribbons. We explore the characteristics
of the electronic band structure with a focus on the nature of edge
states. Edge reconstruction allows the appearance of a new type of
egde states. They are dispersive, with non-zero amplitudes in both
sub-lattices; furthermore, the amplitudes have two components that
decrease with different decay lengths with the distance from the edge;
at the Dirac points one of these lengths diverges, whereas the other
remains finite, of the order of the lattice parameter. We trace this
curious effect to the doubling of the unit cell along the edge, brought
about by the edge reconstruction. In the presence of a magnetic field,
the zero-energy Landau level is no longer degenerate with edge states
as in the case of pristine zigzag ribbon. 
\end{abstract}
\maketitle

\section{Introduction}

At the present time, the most promising scalable growth methods of
graphene films are either based on epitaxial growth on silicon carbide
\cite{Berger_JPCB:2004,DeHeerMRS} or on chemical vapor deposition
(CVD) of graphene on metal surfaces.\cite{Ruoff,MITCVD,Samsung,Rollto}
Yet, the latter methods do not produce graphene films with electronic
mobilities as high as those reported for exfoliated graphene. \cite{Novoselov_Science:2004,PNAS}
Electronic transport \cite{Neto_RMP:2009,Peres_RMP:2010} in CVD-grown
graphene is hindered by grains, grain boundaries and atomic patchwork
quilts,\cite{topological,Incze_APL:2011} which can be interpreted
as topological defects.\cite{topologicalTheory,Banhart_ACSNano:2011}

CVD-grown materials are in general polycrystalline in nature, having
their physical properties dominated by the grain boundaries' size.
The situation is no different for graphene. For this material, it
is theoretically expected that some of its electronic properties will
be markedly different from its exfoliated counterpart, as suggested
by calculations of formation energies of different types of grain
boundaries\cite{EnergyTopolo} and by the transport measurements and
theoretical calculations\cite{AiresEPL} in high-quality CVD-grown\cite{Rollto}
graphene.

Due to graphene's hexagonal structure, originated from the sp$^{2}$
bonds, the grain boundaries are expected to be formed of pentagons-heptagons
pairs, known as Stone-Wales (SW) defects.\cite{Stone_Science:1986}
Recent atomic resolution TEM studies\cite{Meyer_NL:2008,MetallicWire,topological}
have allowed to visualize grain boundaries in CVD-grown graphene.
These experimental studies have shown that the grain boundaries are
not perfectly straight lines and that the 5-7 defects along the boundaries
are not periodic. These type of defects have a profound effect on
the threshold for mechanical failure of the graphene membranes, which
is reduced by an order of magnitude, relatively to the exfoliated
membranes. In what concerns the electronic properties, it has been
shown that the measured electronic mobilities depend on the details
of the CVD-growth recipes. \cite{Ruoff,MITCVD,Rollto,topological}

Furthermore, as shown by recent TEM studies,\cite{topological} these
extended pentagons-heptagons pairs defect lines intercept each other
at random angles, forming irregular polygons with edges showing stochastic
distribution of length, making it extremely difficult to make theoretical
studies of these defects using microscopic tight-binding models. On
a different tone, it has been argued that these defect lines can act
as one-dimensional conducting charged wires.\cite{MetallicWire,Bahamon_PRB:2011}
The charging of these topological wires is achieved by the self-doping
mechanism. \cite{PRB06}

As said, studying Stone-Wales defects in the bulk of graphene, using
microscopic tight-binding models, is a difficult task, due to the
breaking of translational geometry. On the other hand, the grain boundary
formed by the 5-7 defect lines effectively create an edge, giving
rise to an enhanced density of states\cite{MetallicWire,Bahamon_PRB:2011}
at the Dirac point, all in all equal to what is found at the edges
of zigzag nanoribbons. \cite{Nakada_PRB:1996,Fujita_JPSJ:1996,Wakabayashi_PRB:1999,Wakabayashi_Review}
Evading the difficulty of studying topological defects in the bulk
of graphene, we take, in this article, the approach of studying the
formation of Stone-Wales defects at the edges of zigzag nanoribbons,
supported by the experimental findings that grain boundaries effectively
act as edges of the crystalline grain.\cite{MetallicWire,Bahamon_PRB:2011}
We will be focusing our study on the electronic properties of graphene
nanoribbons close to the Dirac point, for the effect of Stone-Wales
defects have their largest impact on the properties of graphene at
low energies.

\textit{Ab-initio} calculations have shown that when SW defects are
present in graphene nanoribbons (GNRs), the energy decreases as the
defect gets closer to the edge of the ribbon.\cite{Huang_PRL:2009}
Other first principles studies have shown that the formation of SW
defects at the edges of both armchair and zigzag GNRs (respectively,
AGNRs and ZGNRs), stabilize them both energetically and mechanically.\cite{Huang_PRL:2009,Koskinen_PRL:2008,Bhowmick_PRB:2010}
The zigzag edge, in particular, is metastable under total reconstruction
with SW defects, and a planar reconstruction spontaneously takes place
at room temperature.\cite{Koskinen_PRL:2008,Lee_PRB:2010}

Edge-reconstructed ZGNRs by means of SW defects, are claimed to be
stable only at very low hydrogen pressure (well below the hydrogen
pressure at ambient conditions) and very low temperatures.\cite{Wassmann_PRL:2008}
However, reconstructions of the zigzag (as well as armchair) edges
have been recently observed with high-resolution TEM.\cite{Koskinen_PRB:2009,Girit_Science:2009,Chuvilin_NJP:2009}
The recent work of Suenaga \textit{et al.},\cite{Suenaga_Nature:2010}
on single-atom spectroscopy using low-voltage STEM, may be used as
yet another mean of identifying edge reconstructions of graphene ribbons.
Moreover, refinements in other techniques, such as Raman spectra of
the edges,\cite{Malola_EPJD:2009} STM images of the edges,\cite{Koskinen_PRL:2008}
and coherent electron focusing,\cite{Rakyta_PRB:2010} may help in
the identification of these kinds of edge reconstructions.

\begin{figure}[htp!]
\includegraphics[width=0.98\columnwidth]{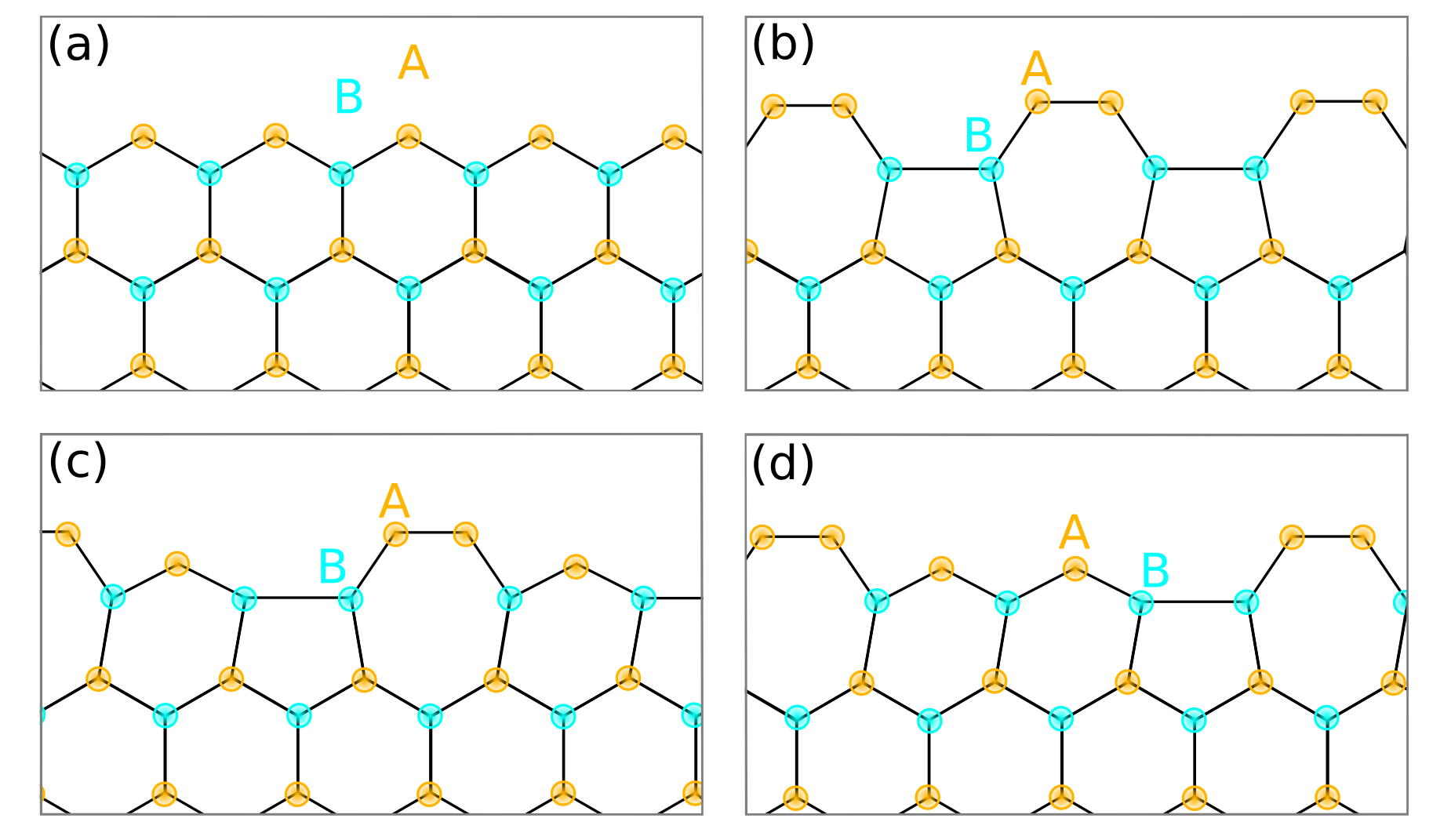} \caption{SW-reconstructed zigzag edges: the pristine zigzag edge in $(a)$,
also known as $zz$ edge; the $zz(57)$ edge in $(b)$; the $zz(576)$
in $(c)$; the $zz(5766)$ in $(d)$;}

\label{fig:reczag_family} 
\end{figure}

In this work, we have studied various reconstructions of zigzag edges
with SW defects, namely $zz(57)$, $zz(576)$ and $zz(5766)$ (see
Fig. \ref{fig:reczag_family}). However, in this article, we give
special emphasis to the case of total reconstruction of the zigzag
edges, $zz(57)$, because it is the most stable configuration in the
absence of hydrogen passivation.\cite{Wassmann_PRL:2008,Koskinen_PRL:2008,Huang_PRL:2009}

This article is organized as follows: In section \ref{sec:TB_edgeRecStudy},
we study the electronic structure of wide zigzag ribbons, with edges
reconstructed by Stone-Wales defects, using an empirical tight-binding
model. In Subsection \ref{subsec:TB_parametrization}, we start by
computing the model parameters using the results of \textit{ab-initio}
simulations. Based on the empirical tight-binding model presented
in Subsection \ref{sec:TBHam}, we study the electronic structure
of ZGNRs whose edges were reconstructed by SW defects, with a focus
on the edge states showing up in these systems (Subsection \ref{subsec:TB_edgeStates}).
We find that some modifications (relatively to the pristine ZGNR)
are introduced in the electronic structure, as well as, that the edge
states of the edge reconstructed ZGNRs are distinct from those of
the pristine ZGNRs. Finally, in Subsection \ref{subsec:TB_landau},
we explore the implications of the presence of a magnetic field directed
perpendicularly to the ribbon plane, in the electronic structure and
edge states of a zigzag ribbon, pinpointing the modifications originating
from the edge reconstruction.


\section{Tight-binding study of ribbons with reconstructed edges}

\label{sec:TB_edgeRecStudy}

In this Section, the main issue under discussion is the behavior of
the edge states of wide zigzag ribbons, whose edges have been reconstructed
due to the formation of edge Stone-Wales defects (see Fig. \ref{fig:reczag_family}).
The huge amount of computational resources needed to study large physical
systems employing \textit{ab-initio} methods, makes it prohibitive
to explore the physics of wide ribbons using such techniques. An alternative
approach, is to use phenomenological tight-binding models, which being
computationally not so demanding also give a microscopic understanding
of the electronic properties of these types of systems.

In order to provide an accurate (tight-binding) description of the
reconstructed edges, we start by performing \textit{ab-initio} simulations
of narrow ribbons, from which we extract the values of the hopping
amplitudes at the edge. These hopping amplitudes are posteriorly used
in the construction of the tight-binding Hamiltonian, in which the
study of the edge states for large ribbons is based on.


\subsection{Parametrization of the hopping amplitudes using \textit{ab-initio}
methods}

\label{subsec:TB_parametrization}

We used Density Functional Theory (DFT) to parametrize the tight-binding.
The calculations were performed using the code \textsc{aimpro},\cite{Rayson_CPC:2008}
under the Local Density Approximation. The Brillouin-zone (BZ) was
sampled for integrations according to the scheme proposed by Monkhorst-Pack.\cite{Monkhorst_PRB:1976}
The core states were accounted for by using the dual-space separable
pseudo-potentials by Hartwigsen, Goedecker, and Hutter.\cite{Hartwigsen_PRB:1998}
The valence states were expanded over a set of $s$-, $p$-, and $d$-like
Cartesian-Gaussian Bloch atom-centered functions. The \textbf{k}-point
sampling was $16\times2\times1$ and the atoms were relaxed in order
to find the equilibrium positions. A supercell with orthorhombic symmetry
was used; the cell parameter in the infinite direction was $4.885\mathring{A}$.
A vacuum layer of $12.7\mathring{A}$ in the nanoribbon plane and
$10.6\mathring{A}$ in the normal direction were used in order to
avoid interactions between nanoribbons in different unit cells.

In what follows, we will focus on the ZGNR with totally reconstructed
edges, the most stable of this family of reconstructions in the absence
of hydrogen passivation (see Fig. \ref{fig:reczag_family}).\cite{Wassmann_PRL:2008,Koskinen_PRL:2008,Huang_PRL:2009}
Note that the dangling bonds that are on the origin of the zigzag
edge reactiveness, are eliminated by the reconstruction of the edge,
forming triple bonds between the outer carbon atoms at the edges ($h_{2}$
bond in Fig. \ref{fig:zz(57)_details}). In the literature, the SW
totally reconstructed edge is usually named as $zz(57)$. Note that
the unit cell of such a ZGNR has twice the size of the unit cell of
the pristine ZGNR (see Fig. \ref{fig:zz(57)}). The generalization
of the following study for SW edge reconstructions, other than $zz(57)$,
for example $zz(576)$, $zz(5766)$, etc., is straightforward.

\begin{figure}[htp!]
\centering \includegraphics[width=0.98\columnwidth]{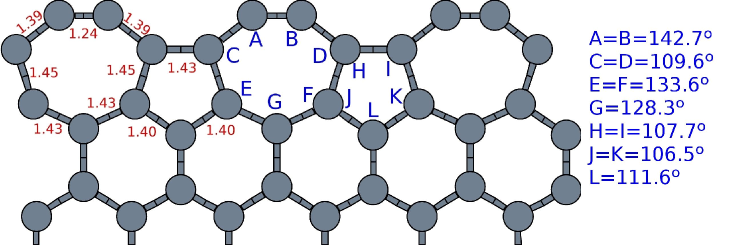}
\caption{Relaxed edge geometry of the totally reconstructed zigzag edge (in
absence of hydrogen passivation), $zz(57)$, obtained from the \textit{ab-initio}
calculations. The numbers refer to the bond lengths in angstroms;
the capital letters refer to the angles between two adjacent bonds.}

\label{fig:SW57_parameters} 
\end{figure}

In Fig. \ref{fig:SW57_parameters}, we show the relaxed edge geometry
of a totally reconstructed edge (in absence of hydrogen passivation),
$zz(57)$, as obtained from the \textit{ab-initio} calculations, together
with the inter-carbon distances and the angles between carbon bonds.

The procedure for determining the hopping amplitudes at the reconstructed
edge is the following. From \textit{ab-initio} calculations one obtains
the different carbon-carbon distances at the edges of the ribbons
as well as the values of the angles between carbon bonds (see Fig.
\ref{fig:SW57_parameters}). In the case of the $zz(57)$ edge reconstruction,
our first principles calculations show that the ribbons remain planar
(we have allowed the system to relax along the three spatial dimensions);
therefore the values of the angles in Fig. \ref{fig:SW57_parameters}
play no role in the determination of the hopping amplitudes, since
these arise from $pp\pi$ hybridization. Using the carbon-carbon distances,
we compute the hopping amplitudes using the parametrization\cite{Tang_PRB:1996}
\begin{eqnarray}
\tau(r_{ij})=\big(\frac{r_{ij}}{a_{0}}\big)^{-\alpha_{2}}\exp[-\alpha_{3}\times(r_{ij}^{\alpha_{4}}-a_{0}^{\alpha_{4}})],\label{eq:hoppParam}
\end{eqnarray}
where $r_{ij}$ stands for the distance between the carbons labeled
by $i$ and $j$ (given in unities of Angstr\"oms), the adimensional
parameters $\alpha_{2}=1.2785$, $\alpha_{3}=0.1383$, $\alpha_{4}=3.4490$,
while $a_{0}$ is the carbon-carbon distance in the bulk (in units
of Angstr\"oms).\cite{Tang_PRB:1996} The hopping renormalizations,
$h_{i}$, $h_{i}'$ and $v_{i}$, (see Fig. \ref{fig:zz(57)_details}
for their definition) are given by the $\tau(r_{ij})$ for the corresponding
carbon-carbon distances at the edge. For the $zz(57)$ edge, in the
absence of passivation, the values of these renormalizations are listed
in Table \ref{tab_hopp}.

\begin{table}
\begin{tabular}{cccccccccccc}
\hline 
$h_{1}$  & $h_{2}$  & $h_{3}$  & $h_{4}$  & $h_{1}'$  & $h_{2}'$  & $h_{3}'$  & $h_{4}'$  & $v_{1}$  & $v_{2}$  &  & \tabularnewline
\hline 
\hline 
1.06  & 1.42  & 1.06  & 0.94  & 0.98  & 0.98  & 1.04  & 1.04  & 0.94  & 0.94  &  & \tabularnewline
\hline 
\end{tabular}\caption{Values of the hoppings in unities of $t$ (which we also call hopping
renormalizations) for a $zz(57)$ (see Fig. \ref{fig:SW57_parameters}
and Fig. \ref{fig:zz(57)_details}) calculated from the C-C distances
obtained from the DFT numerical calculations using Eq. (\ref{eq:hoppParam}).}

\label{tab_hopp} 
\end{table}


\subsection{The Tight-Binding Hamiltonian of a ZGNR with $zz(57)$ edges}

\label{sec:TBHam}

The simplest model one can construct describing non-interacting electrons
in a ZGNR whose edges have been reconstructed by SW defects is the
first neighbor tight-binding (TB) model. Generically, a ribbon has
$N$ zigzag rows of atoms along the longitudinal direction ($0\leq n\leq N-1$)
and, in each unit cell, there are $P$ zigzag columns of atoms ($1\leq p\leq P$).
In the case of a $zz(57)$ edge, $P=2$.

\begin{figure}[htp!]
\includegraphics[width=0.98\columnwidth]{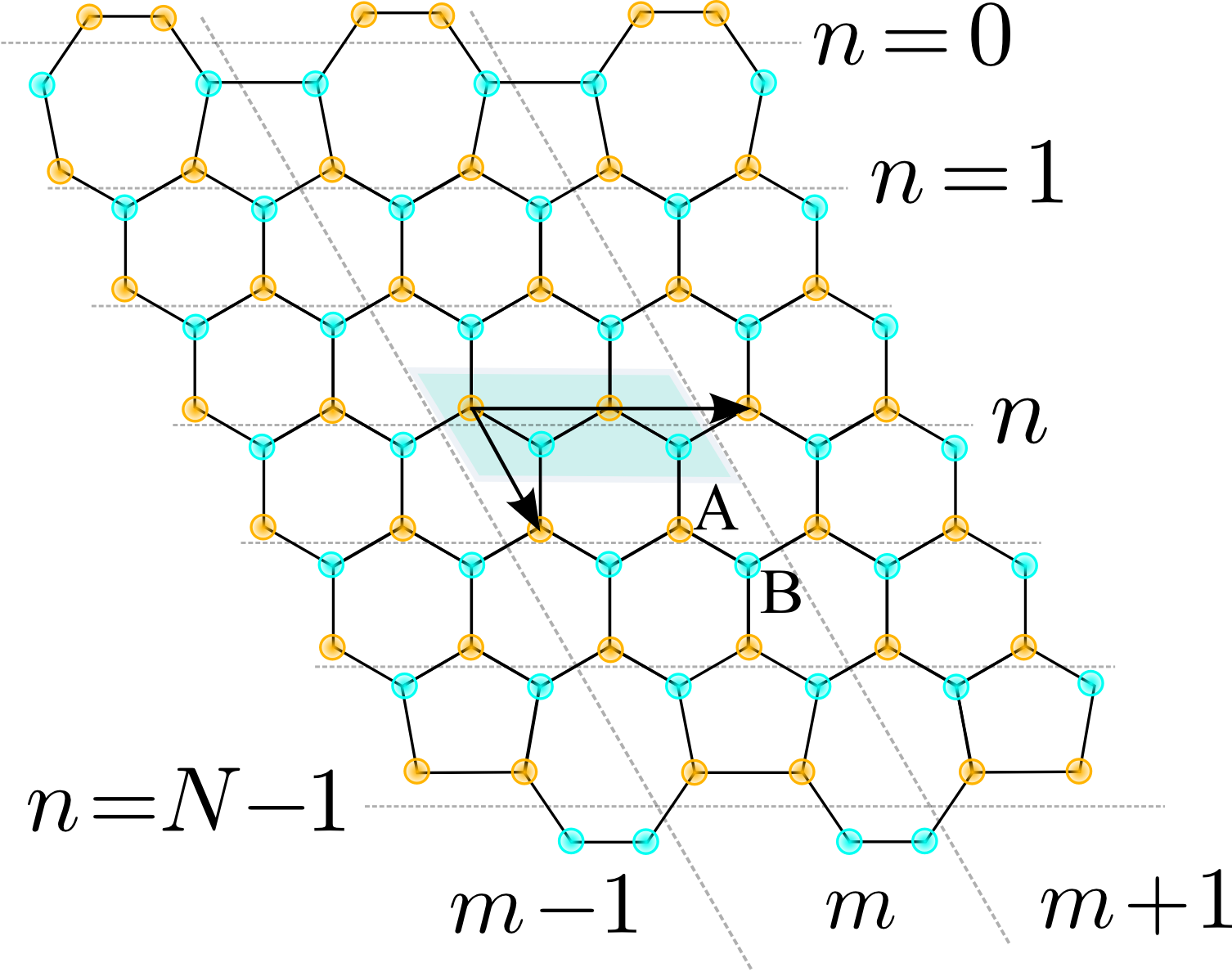} \caption{Scheme of a ZGNR with its edges totally reconstructed by SW defects
{[}a $zz$(57) ribbon{]}. Details of the edge are represented in Fig.
\ref{fig:zz(57)_details}.}

\label{fig:zz(57)} 
\end{figure}

The TB Hamiltonian for the edge reconstructed ZGNR, can be written
as 
\begin{eqnarray}
H & = & H^{U}+H^{bulk}+H^{L},\label{eq:Hamiltonian}
\end{eqnarray}
where $H^{U}$ stands for the Hamiltonian of the region in the vicinity
of the upper edge of the ribbon (at $n=0$ in Fig. \ref{fig:zz(57)}),
$H^{L}$ stands for the region in the vicinity of the lower edge (at
$n=N-1$ in Fig. \ref{fig:zz(57)}) and $H^{bulk}$ stands for the
bulk of the ribbon.

\begin{figure}[htp!]
\includegraphics[width=0.88\columnwidth]{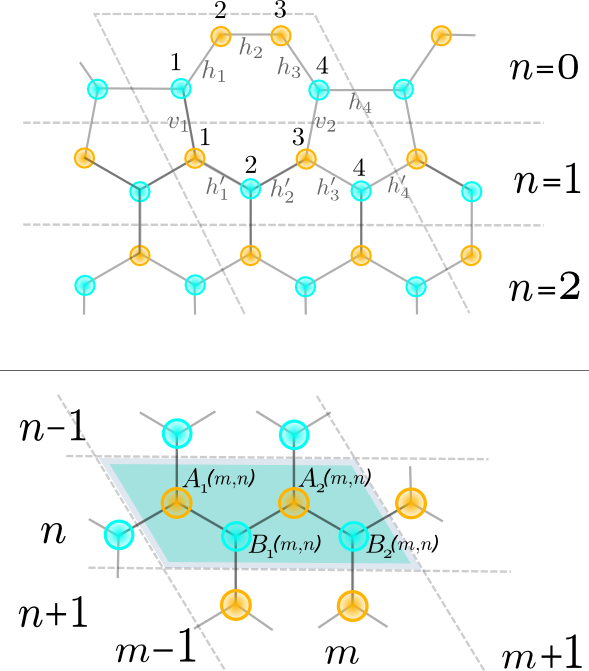}
\caption{Details of the ZGNR with its edges totally reconstructed by SW defects,
$zz(57)$. The $h_{i}$, $h_{i}'$ and $v_{i}$ stand for the factors
giving the renormalization of the hoppings, $t$, between nearest
neighbors in the vicinity of the defect. We omit the lower edge because
it is analogous to the upper one.}

\label{fig:zz(57)_details} 
\end{figure}

The \textit{ab-initio} results (see Fig. \ref{fig:SW57_parameters})
show that only in the first two rows are the hopping parameters between
two adjacent carbon atoms different from their usual value in the
bulk. Thus, we choose to identify term $H^{U}$ ($H^{L}$) in the
full Hamiltonian with the two upper (lower) rows of atoms of the ribbon.
The annihilation operators of the four numbered atoms in row $n=0$
(see Fig. \ref{fig:zz(57)_details}), are denoted by $d_{1}(m)$,
$d_{2}(m)$, $d_{3}(m)$, and $d_{4}(m)$, while those referring to
the four numbered atoms in row $n=1$, are denoted by $c_{1}(m)$,
$c_{2}(m)$, $c_{3}(m)$, and $c_{4}(m)$.

For the sake of clarity, we will separate in $H^{U}$ the terms referring
to each row, $n=0$ and $n=1$, and to the coupling between them,
$H^{U}=H_{n=0}^{U}+H_{n=1}^{U}+H_{c}^{U}$. For row $n=0$, we have
\begin{eqnarray}
H_{n=0}^{U} & = & -t\sum_{m}\bigg\{\sum_{i=1}^{3}\Big[h_{i}d_{i}^{\dagger}(m)d_{i+1}(m)\Big]\nonumber \\
 & + & h_{4}d_{4}^{\dagger}(m)d_{1}(m+1)+{\rm h.c.}\bigg\},\label{eq:Hupper0}
\end{eqnarray}
 while for row $n=1$, 
\begin{eqnarray}
H_{n=1}^{U} & = & -t\sum_{m}\bigg\{\sum_{i=1}^{3}\Big[h_{i}'c_{i}^{\dagger}(m)c_{i+1}(m)\Big]\nonumber \\
 & + & h_{4}'c_{4}^{\dagger}(m)c_{1}(m+1)\Big]+{\rm h.c.}\bigg\},\label{eq:Hupper1}
\end{eqnarray}
 and for the coupling between row $n=0$ and row $n=1$, 
\begin{eqnarray}
H_{c}^{U} & = & -t\sum_{m}\Big[v_{1}c_{1}^{\dagger}(m)d_{1}(m)+v_{2}c_{3}^{\dagger}(m)d_{4}(m)\Big]+{\rm h.c.}.\nonumber \\
\label{eq:HupperCoup}
\end{eqnarray}
 Recall from Table \ref{tab_hopp} that $h_{1}=h_{3}$, $h_{1}'=h_{2}'$,
$h_{3}'=h_{4}'$, and $v_{1}=v_{2}$.

The term $H^{bulk}$, corresponding to the Hamiltonian of the bulk
(between row $n=2$ and $n=N-3$), is given by 
\begin{eqnarray}
H^{bulk} & = & -t\sum_{m}\sum_{n=2}^{N-3}\bigg(\Big[a_{1}^{\dagger}(m;n)+a_{2}^{\dagger}(m;n)\nonumber \\
 & + & a_{1}^{\dagger}(m;n+1)\Big]b_{1}(m;n)+\Big[a_{2}^{\dagger}(m;n)\nonumber \\
 & + & a_{1}^{\dagger}(m+1;n)+a_{2}^{\dagger}(m;n+1)\Big]b_{2}(m;n)+{\rm h.c.}\bigg),\nonumber \\
\label{eq:Hbulk}
\end{eqnarray}
 where $a_{p}(m;n)$ {[}$b_{p}(m;n)${]} is the annihilation operator
of an electron state localized at the atom of sub-lattice $A$ ($B$)
in column $p$ ($p=1,\,2$ for a $zz(57)$ edge) and row $n$, of
the unit cell labeled by $m$. The term $H^{L}$ describing the lower
edge is analogous to the upper edge term, $H^{U}$. Recall that the
$h_{i}$, $v$ and $h_{i}'$ parameters in the equations defining
the tight-binding Hamiltonian correspond to the values of the hoppings
in units of $t$. In addition, we make the following identifications:
\begin{eqnarray*}
d_{1(4)}(m) & \to & b_{1(2)}(m;0),\\
d_{2(3)}(m) & \to & a_{1(2)}(m;0),\\
c_{1(3)}(m) & \to & a_{1(2)}(m;1),\\
c_{2(4)}(m) & \to & b_{1(2)}(m;1).
\end{eqnarray*}

With no loss of generality, we assume periodic boundary conditions
along the ribbon $x$-direction. This simplification allows us to
diagonalize the Hamiltonian with respect to $m$ by Fourier transforming
$H$ along the $x$-direction, 
\begin{equation}
H=\sum_{k}H_{k}=\sum_{k}\big[H_{k}^{U}+H_{k}^{bulk}+H_{k}^{L}\big]\,.\label{eq_Hamilt_Fourier}
\end{equation}

\begin{figure}[htp!]
\includegraphics[width=0.98\columnwidth]{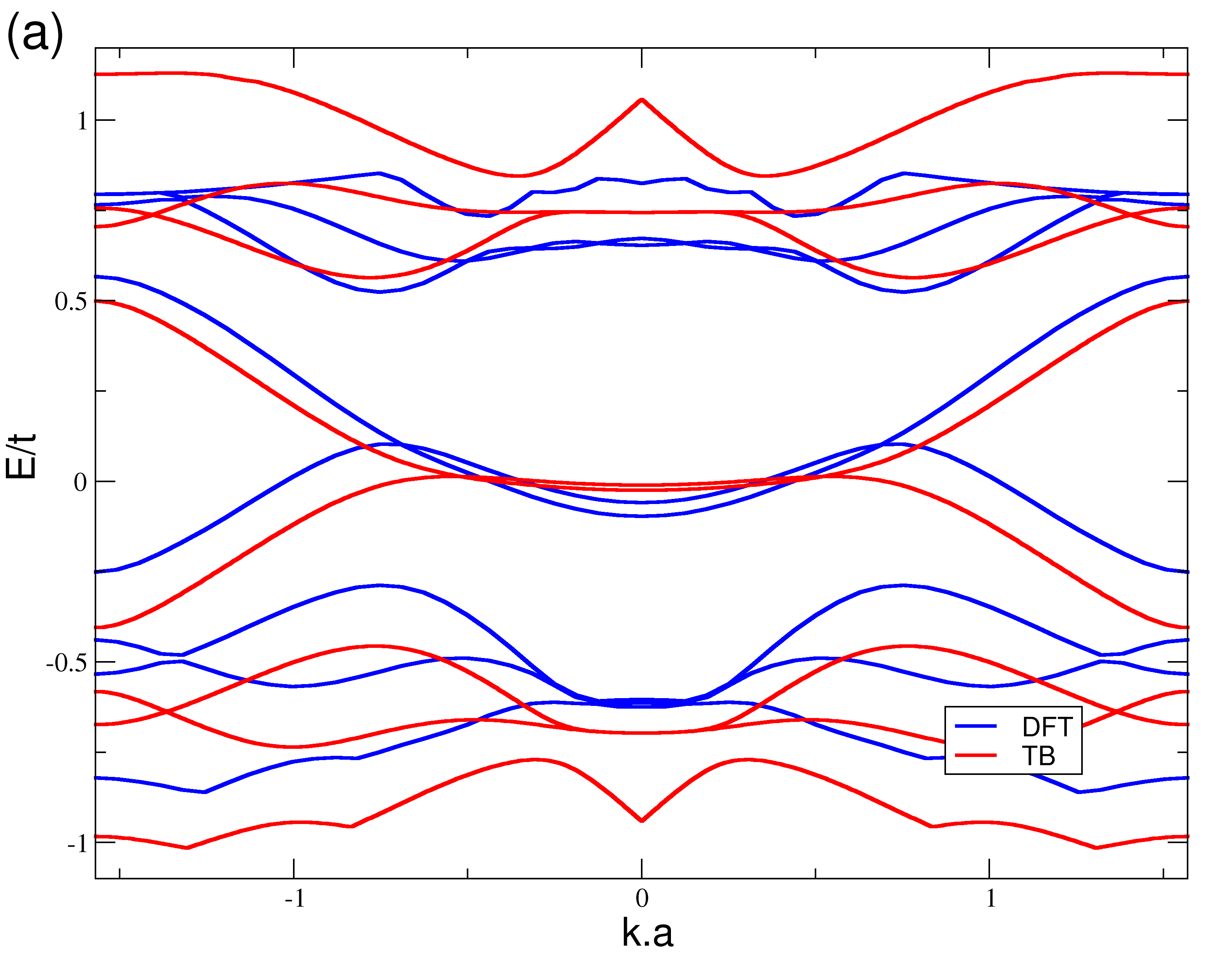}
\includegraphics[width=0.98\columnwidth]{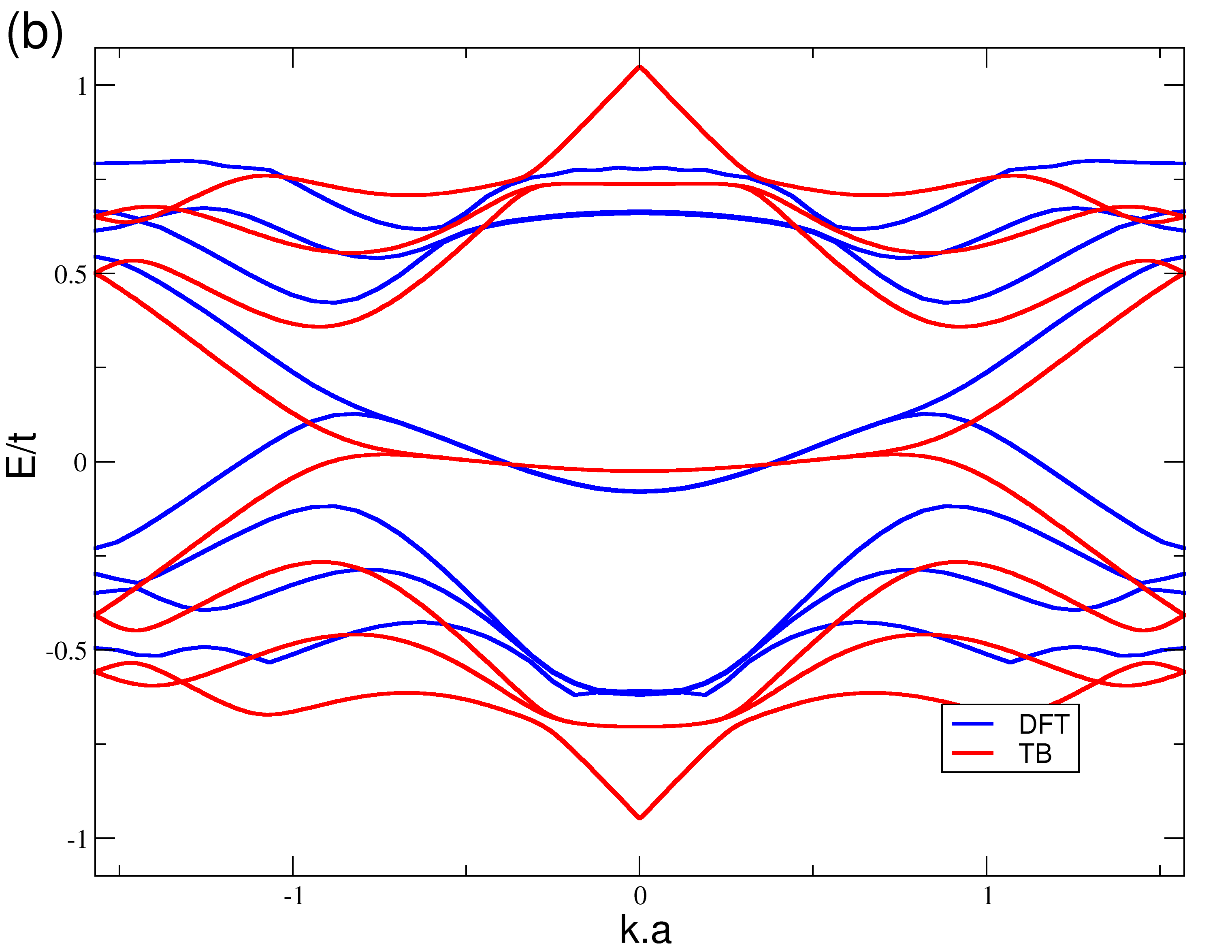}
\caption{Comparison between the low-energy spectrum of a ZGNR (with both edges
reconstructed) obtained with DFT (blue) and TB (red). In $(a)$ the
ribbon has a width of $18\mathring{A}$ (or $N=8$ zigzag lines),
while in $(b)$ the ribbon has a width of $31\mathring{A}$ (or $N=14$
zigzag lines). The Fermi level is at $E/t=0$.}

\label{fig:DFT_TB_SWD-57_TwoSide} 
\end{figure}

Having determined the values of the $h_{i}$, $v$ and $h'_{i}$ (see
Table \ref{tab_hopp}), we compare in Fig. \ref{fig:DFT_TB_SWD-57_TwoSide}
the obtained low-energy spectrum from the \textit{ab-initio} calculations
with that resulting from the numerical diagonalization%
\footnote{The numerical diagonalization of the tight-binding Hamiltonian was
performed using the tools of LAPACK numerical library.%
} of the tight-binding Hamiltonian $H_{k}$, Eq. (\ref{eq_Hamilt_Fourier}).

As we can see in Fig. \ref{fig:DFT_TB_SWD-57_TwoSide}, the DFT and
TB numerical calculations for narrow zigzag ribbons with $zz(57)$
edges originate low-energy spectra with similar features. The differences
between the DFT and the TB spectra are probably due both to the simplified
character of the TB treatment (especially the first-neighbor approximation)
and to finite size effects affecting both systems differently. In
fact, it is well known that even for an accurate description of \textit{ab-initio}
of bulk graphene bands, one needs a tight-binding model including
hoppings up to third-nearest neighbors.\cite{Ordejon} Since our interest
is the understanding of the main features of the low-energy spectra,
we keep in the tight-binding model only the first neighbor hopping.

In the reduced Brillouin zone, arising from the doubling of the unit
cell along the edge ($x$) direction, the Dirac points of bulk graphene
appear at $\mathbf{K}=2\pi(1/2,-\sqrt{3}/2)/3$ and $\mathbf{K}'=2\pi(-1/2,\sqrt{3}/2)/3$.
In a ribbon, they will show up at $ka=\pi/3$ and at $ka=-\pi/3$,
where $k$ is the momentum along the edge direction. We now focus
on the dispersive energy levels present around the Fermi level, appearing
between these two Dirac points.


\subsection{Edge states of a $zz(57)$ edge}

\label{subsec:TB_edgeStates}

\begin{figure}[htp!]
\includegraphics[width=0.48\columnwidth]{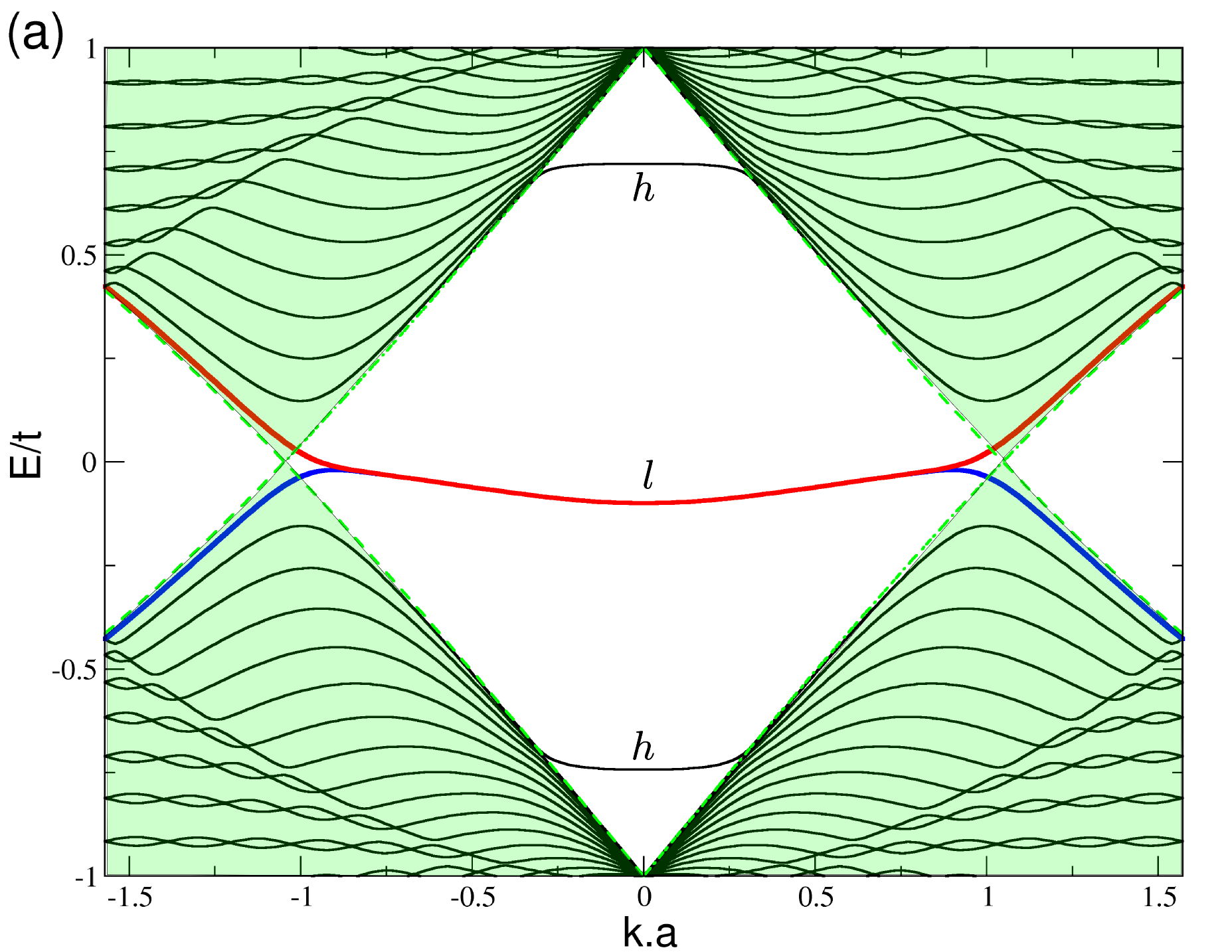}
\includegraphics[width=0.48\columnwidth]{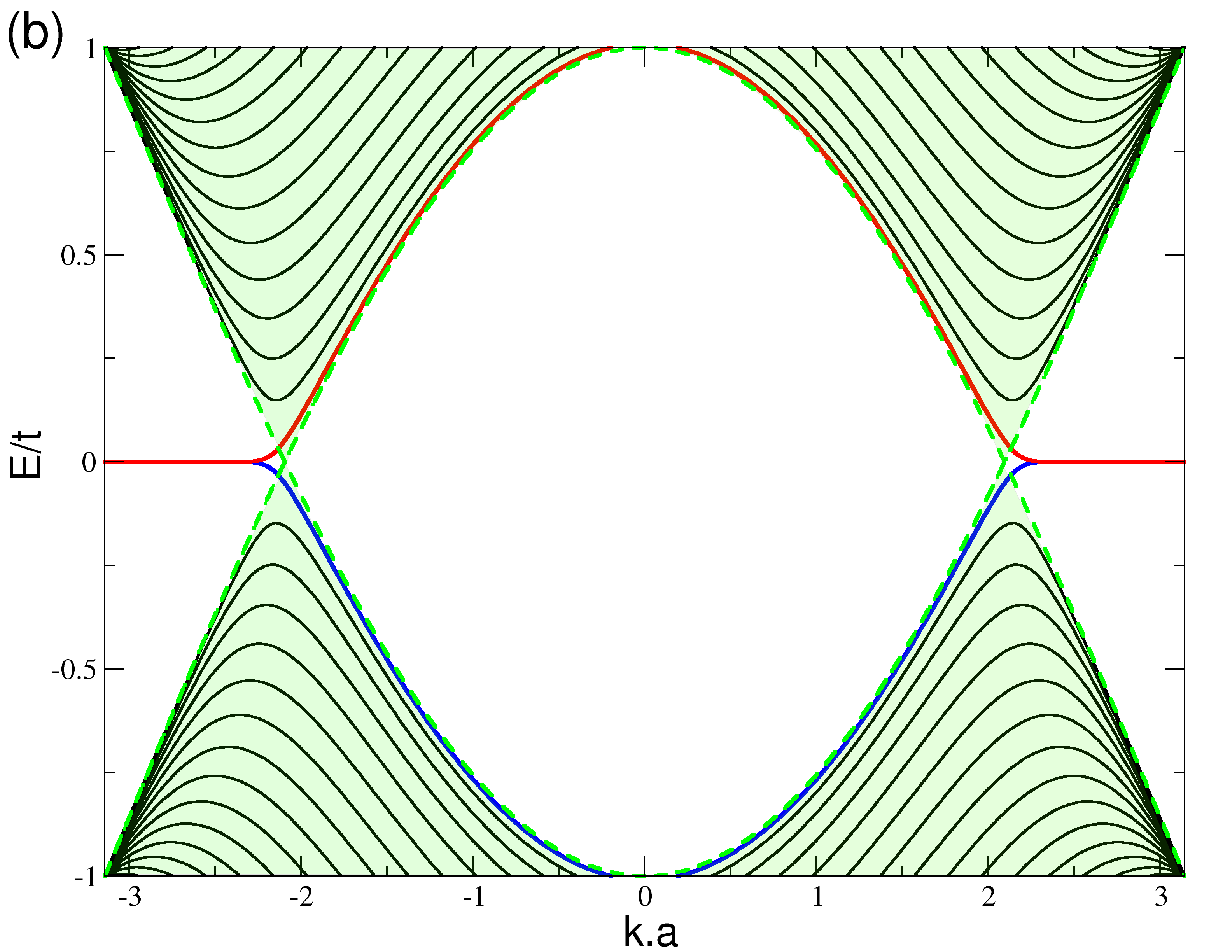}
\caption{$(a)$: Low-energy spectrum of a ZGNR with $N=30$ and both edges
totally reconstructed with SW defects, $zz(57)$. Remember that the
energy spectrum of a totally reconstructed edge, results from a doubled
unit cell relatively to the pristine ZGNRs, and consequently is double-folded
relatively to the latter. The labeled energy levels $(b)$: Low-energy
spectrum of a pristine ZGNR with $N=30$ and zigzag edges. Both $(a)$
and $(b)$ were obtained using a tight-binding model. In both cases,
the shaded areas indicate which levels are allowed in bulk graphene.
Eigenstates corresponding to levels that are outside the shaded region,
will necessarily be located at the edges of the ribbon both in $(a)$
$zz(57)$ edges and in $(b)$ perfect zigzag edges. In the $(a)$
panel, the levels labeled by $h$ and $l$, stand, respectively, for
high and low-energy. }

\label{fig:Comparing_Bulk-Ribbon} 
\end{figure}

In a finite graphene sheet, energy levels appearing outside the range
of allowed electronic states of bulk graphene correspond to states
localized at the edges, called `edge states', as usual in surface
physics. Consequently, from Fig. \ref{fig:Comparing_Bulk-Ribbon}$(a)$,
we can guess that $zz(57)$ edges allow both high and low-energy edge
states (respectively, the levels $h$ and $l$ in Fig. \ref{fig:Comparing_Bulk-Ribbon});
this contrasts with what happens in the pristine zigzag edge (only
low-energy edge states).\cite{Fujita_JPSJ:1996,Nakada_PRB:1996,Wakabayashi_PRB:1999}
In what follows, we will focus on the physically relevant low-energy
ones.

Since edge states decay exponentially into the bulk, in wide ribbons
they can be studied as states of semi-infinite ribbons: states at
different edges are uncoupled if the ribbon width is much larger than
the decay length. In the case of a semi-infinite ribbon with pristine
zigzag edges, the edge states occur at zero-energy.\cite{Nakada_PRB:1996,Wakabayashi_PRB:1999}
In such a case, the tight-binding equations simplify to independent
recurrence relations for the amplitudes of the $A$ and $B$ sub-lattices,
which yield, transparently, the exact wave-functions, the analytical
expression of the decay length as function of the momentum along the
edge, and the range of momentum values in which such states are possible.
In the case of a zigzag ribbon with edges totally reconstructed with
SW defects, we face the complication that edge states \emph{have dispersion},
and are not at zero energy.

To investigate the possibility of low energy edge states of such a
system, we must solve the Schr\"odinger equation, 
\begin{eqnarray}
H_{k}|\mu,k\rangle & = & \epsilon_{\mu,k}|\mu,k\rangle\label{eq:generalH}
\end{eqnarray}
 for $\left|\epsilon_{\mu,k}\right|/t\ll1$, where $H_{k}$ is the
same as that obtained from the transformation in Eq. \ref{eq_Hamilt_Fourier},
of the Hamiltonian given by Eqs. (\ref{eq:Hamiltonian})-(\ref{eq:Hbulk})
with $n\geq0$. Note that $H_{k}$ defines effectively a 1D problem
in the transverse direction of the ribbon. Consequently, we can express
any eigenstate $|\mu,k\rangle$ as a linear combination of the site
amplitudes along the transverse direction of the ribbon, 
\begin{eqnarray}
|\mu,k\rangle & = & \sum_{n=0}^{N-1}\sum_{p=1}^{2}\Big[A_{p}(k;n)|a;k;p,n\rangle\nonumber \\
 & + & B_{p}(k;n)|b;k;p,n\rangle\Big],
\end{eqnarray}
 with the one-particle states, $|r;k;p,n\rangle=r_{p}^{\dagger}(k;n)|0\rangle$,
where $p=1,2$ and $n=1,\dots,N-1$ define, respectively, the column
and the line of the unit cell, and $r=a,b$. To lighten our notation,
we have identified the states at the upper edge as \begin{subequations}
\label{eq:NotChangeUp} 
\begin{eqnarray}
|d_{1(4)};k\rangle & = & |b;k;1(2),1\rangle,\label{eq:NotChange1}\\
|d_{2(3)};k\rangle & = & |a;k;1(2),0\rangle,\label{eq:NotChange2}\\
|c_{1(3)};k\rangle & = & |a;k;1(2),1\rangle,\label{eq:NotChange3}\\
|c_{2(4)};k\rangle & = & |b;k;1(2),2\rangle,\label{eq:NotChange4}
\end{eqnarray}
 \end{subequations} while at the lower edge \begin{subequations}
\begin{eqnarray}
|c_{5(7)};k\rangle & = & |a;k;1(2),N-2\rangle,\label{eq:NotChange5}\\
|c_{6(8)};k\rangle & = & |b;k;1(2),N-2\rangle,\label{eq:NotChange6}\\
|d_{5(8)};k\rangle & = & |a;k;1(2),N-1\rangle,\label{eq:NotChange7}\\
|d_{6(7)};k\rangle & = & |b;k;1(2),N-1\rangle.\label{eq:NotChange8}
\end{eqnarray}
 \end{subequations} Note that there are four states per zigzag row
(identified by $n$), coming from the four sub-lattices $A_{1}$,
$B_{1}$, $A_{2}$ and $B_{2}$. Equating coefficients, we obtain
a set of $2\times2\times N$ (tight-binding) equations, where $N$
is the number of zigzag rows of atoms in the unit cell.

To build an analytical description for edge states in a semi-infinite
ribbon, with row index $n\ge0$, we write the TB equations in matrix
form, where $\boldsymbol{A}(k;n)=\big[A_{1}(k;n),A_{2}(k;n)\big]^{T}$
and $\boldsymbol{B}(k;n)=\big[B_{1}(k;n),B_{2}(k;n)\big]^{T}$ will
stand for column vectors.

For rows with $n>1$, the relations between the amplitudes are the
same as those of a pristine ribbon: \begin{subequations}\label{eq:recurrence_bulk}
\begin{align}
\mathbf{A}(k;n+1) & -W_{A}\mathbf{A}(k;n)=-\left(\frac{\epsilon}{t}\right)\mathbf{B}(k;n+1),\label{eq:recurrence_bulk-1-1}\\
\mathbf{B}(k;n) & -W_{B}\mathbf{B}(k;n+1)=-\left(\frac{\epsilon}{t}\right)\mathbf{A}(k;n).\label{eq:recurrence_bulk-2-1}
\end{align}
\end{subequations} The matrices $W_{A}$ and $W_{B}$, defined in
Eqs. (\ref{eq:W_simple}), commute and, therefore, share a common
eigenbasis, $\{\boldsymbol{u^{+}},\boldsymbol{u^{-}}\}$ (see Appendix~\ref{sec:appendix}
for details). Let us denote the corresponding eigenvalues by $\xi_{A}^{\pm}$
and $\xi_{B}^{\pm}$, respectively. These quantities depend on the
value of the longitudinal momentum, $k$ and are given by: \begin{subequations}
\label{eq:xis-1} 
\begin{eqnarray}
\xi_{A}^{+} & = & -2\cos(ka/2)e^{i\frac{ka}{2}},\label{eq:xiA1-1}\\
\xi_{A}^{-} & = & 2i\sin(ka/2)e^{i\frac{ka}{2}},\label{eq:xiA2-1}\\
\xi_{B}^{+} & = & -2\cos(ka/2)e^{-i\frac{ka}{2}}=\left(\xi_{A}^{+}\right)^{*},\label{eq:InvxiB1-1}\\
\xi_{B}^{-} & = & -2i\sin(ka/2)e^{-i\frac{ka}{2}}=\left(\xi_{A}^{-}\right)^{*}.\label{eq:InvxiB2-1}
\end{eqnarray}
 \end{subequations} 

Changing to the $\{\boldsymbol{u^{+}},\boldsymbol{u^{-}}\}$ basis,
\begin{subequations}\label{eq:preDiagRecurr-1} 
\begin{eqnarray}
\boldsymbol{A}(k;n) & = & \alpha_{+}(k;n)\boldsymbol{u^{+}}+\alpha_{-}(k;n)\boldsymbol{u^{-}},\label{eq:preDiagRecurr1-1}\\
\boldsymbol{B}(k;n) & = & \beta_{+}(k;n)\boldsymbol{u^{+}}+\beta_{-}(k;n)\boldsymbol{u^{-}},\label{eq:preDiagRecurr2-1}
\end{eqnarray}
\end{subequations} we obtain Eqs. (\ref{eq:recurrence_bulk}) in
the form, \begin{subequations}\label{eq:tB_diagonal_chain}
\begin{align}
\alpha_{\sigma}(k;n+1)-\xi_{A}^{\sigma}\alpha_{\sigma}(k;n) & =-\left(\frac{\epsilon}{t}\right)\beta_{\sigma}(k;n+1),\label{eq:tB_diagonal_chain_A}\\
\beta_{\sigma}(k;n)-(\xi_{A}^{\sigma})^{*}\beta(k;n+1) & =-\left(\frac{\epsilon}{t}\right)\alpha_{\sigma}(k;n),\label{eq:tB_diagonal_chain_B}
\end{align}
\end{subequations}where $\sigma=\pm1$. Note that with the two possible
values for $\sigma$, Eqs. (\ref{eq:tB_diagonal_chain}) give four
equations. These equations describe two independent 1D $AB$ chains
in the $n$ coordinate, one for each of the modes $\mathbf{u}_{+}$
and $\mathbf{u}_{-}$; the hopping amplitude alternates between $-t$
and $t\xi_{A}^{\sigma}$. 

The two modes $\mathbf{u}_{+}$ and $\mathbf{u}_{-}$ are easily interpreted.
If we look for propagating solutions ($q_{\sigma}$ real), \begin{subequations}
\begin{align}
\alpha_{\sigma}(k;n) & =\alpha_{\sigma}(k)e^{iq_{\sigma}n},\\
\beta_{\sigma}(k;n) & =\beta_{\sigma}(k)e^{iq_{\sigma}n},
\end{align}
\end{subequations} we arrive at the equation 
\begin{equation}
\left(\frac{\epsilon}{t}\right)^{2}=\left|\left(1-e^{-iq_{\sigma}}\xi_{A}^{\sigma}\right)\right|^{2}.\label{eq:Energy_expr_E0}
\end{equation}
Low energy states correspond to $(\epsilon/t)^{2}\ll1$; but it can
easily be checked from Eqs. (\ref{eq:xis-1}), that $\left|\xi_{A}^{+}\right|\ge\sqrt{2}$,
for all $ka$ in the F.B.Z., whereas $\left|\xi_{A}^{-}\right|\approx1$
around $ka=\pm\pi/3$. Hence, propagating states of the $\sigma=+$
modes have an energy of order $t$; the $\sigma=-$ modes are the
low energy bulk states when $k$ is near the Dirac value. The existence
of these two modes reflects the folding of the Brillouin zone to account
for the doubling of the unit cell. At the Bloch momentum of a Dirac
point there are two different energy levels, only one of which is
of low energy, and corresponds to the $\mathbf{u}_{-}$ mode. In fact,
inspecting the relation between the $A_{1}$ and $A_{2}$ amplitudes
in the $\mathbf{u}_{-}$mode {[}see Appendix~\ref{sec:appendix},
Eqs. (\ref{eq:u_pm}){]} one sees that it corresponds to what is expected
from a plane wave at a Dirac point. 

Nevertheless, for decaying states $(q_{\sigma}$ with an imaginary
part), we cannot exclude the possibility that low energy states can
have a $\sigma=+$ component, because in that case, the right hand
side of Eq. (\ref{eq:Energy_expr_E0}) has a factor $(1-e^{-\Im q_{+}}(\xi_{A}^{+})^{*}e^{i\Re q_{+}})$,
which can be close to zero. We will see in a moment that the boundary
conditions (BCs) arising from the $zz(57)$ edge bring about precisely
this situation.

Let us now discuss what kind of solutions are obtained from Eqs. (\ref{eq:tB_diagonal_chain})
if the system supports zero energy states. For zero energy, the bulk
Eqs. (\ref{eq:tB_diagonal_chain}) become independent recursion relations
\begin{subequations}\label{eq:zero_energy_sols}
\begin{eqnarray}
\alpha_{\sigma}(k;n+1) & = & \xi_{A}^{\sigma}\alpha_{\sigma}(k;n),\\
\beta_{\sigma}(k;n+1) & = & \frac{1}{\left(\xi_{A}^{\sigma}\right)^{*}}\beta_{\sigma}(k;n).
\end{eqnarray}
\end{subequations} From Eqs. (\ref{eq:xis-1}), $\left|\xi_{A}^{+}\right|>\sqrt{2}$,
thus requiring $\alpha_{+}(k,n)=0$, otherwise we would have a non-normalizable
state. Also, we must have either $\alpha_{-}(k,n)$ or $\beta_{-}(k,n)=0$,
depending on whether $\left|\xi_{A}^{-}\right|$ is greater or smaller
than $1.$ Consider, for instance, the latter case: the required conditions
for zero energy states would then be $\alpha_{+}(k,n)=\beta_{-}(k,n)=0$. 

The previous paragraph did not impose any type of conditions arising
from the boundary. It turns out that the existence of zero energy
states depends on the specific form of the boundary conditions. We
note however, that in this type of edge reconstruction surface states
always exist, but not necessarily at zero energy. The boundary conditions
can be derived form the tight-binding equations for the rows $n=0,1$.
As shown in Appendix~\ref{sec:appendix}, Eq. (\ref{eq:bc_eigenbasis-1}),
they can be approximated by the zero energy BCs, $\boldsymbol{\alpha}(k;2)=\mathcal{M}\mathbf{\boldsymbol{\beta}}(k;2)$,
where $\mathcal{M}$ is a $k$ dependent matrix defined explicitly
in Appendix~\ref{sec:appendix}; in full, \begin{subequations}\label{eq:bc_main_text}
\begin{align}
\alpha_{+}(k;2) & =\mathcal{M}_{++}\beta_{+}(k;2)+\mathcal{M}_{+-}\beta_{-}(k;2),\\
\alpha_{-}(k;2) & =\mathcal{M}_{-+}\beta_{+}(k;2)+\mathcal{M}_{--}\beta_{-}(k;2).
\end{align}
\end{subequations} In the case where zero energy states exist, the
boundary conditions defined by Eqs. (\ref{eq:bc_main_text}) are exact.
For the $k$ values for which $\left|\xi_{A}^{-}\right|<1$, zero
energy states require, as we have seen, $\alpha_{+}(k)=\beta_{-}(k)=0$;
this is possible only if $\mathcal{M}_{++}=0$. This condition is,
in fact, verified in certain limits, the simplest one corresponding
to ignoring the hopping renormalizations at the edge, that is, taking
$h_{i}=v=h_{i}'=1$, in which case the matrix $\mathcal{M}$ reads
\begin{equation}
\mathcal{M}=-4\sin^{2}(ka)\left[\begin{array}{cc}
0 & (\xi_{A}^{-})^{*}\\
(\xi_{A}^{+})^{*} & 0
\end{array}\right].
\end{equation}
Another interesting limit to consider is $h'_{i}=1.$ In this case,
one obtains
\begin{equation}
\mathcal{M}_{++}\propto h_{1}^{2}-h_{2}h_{4},
\end{equation}
and consequently, zero-energy states should be observed if $h_{1}^{2}-h_{2}h_{4}=0.$ 

We have confirmed these results by numerical diagonalization of the
tight-binding Hamiltonian.%
\footnote{The alternative possibility for zero energy states in the range where
$\left|\xi_{A}^{-}\right|>1$, and $\alpha_{+}(k)=\alpha_{-}(k)=0$,
requires $\det\left[\mathcal{M}(k)\right]=0$; we found no relevant
limits where this is verified.%
} In both situations, as $\mathcal{M}_{++}=0$, the zero-energy states
appear in the range where $\left|\xi_{A}^{-}\right|<1$, \emph{i.e.,}
$\left|ka\right|<\pi/3$, and have the form (for $n>1$)\begin{subequations}\label{eq:wfunctions_zero_energy}
\begin{eqnarray}
\mbox{\ensuremath{\alpha}}_{-}(k;n) & = & \left(\xi_{A}^{-}\right)^{n-2}\alpha_{-}(k),\label{eq:wfunctions_zero_energy-a}\\
\beta_{+}(k;n) & = & \left(\frac{1}{\left(\xi_{A}^{+}\right)^{*}}\right)^{n-2}\beta_{+}(k),\label{eq:wfunctions_zero_energy-b}
\end{eqnarray}
\end{subequations}with 
\begin{equation}
\alpha_{-}(k)=-4\sin^{2}(ka)(\xi_{A}^{+})^{*}\beta_{-}(k).\label{eq:wfunctions_zero_energy-c}
\end{equation}

\begin{figure}[htp!]
\includegraphics[width=0.49\textwidth]{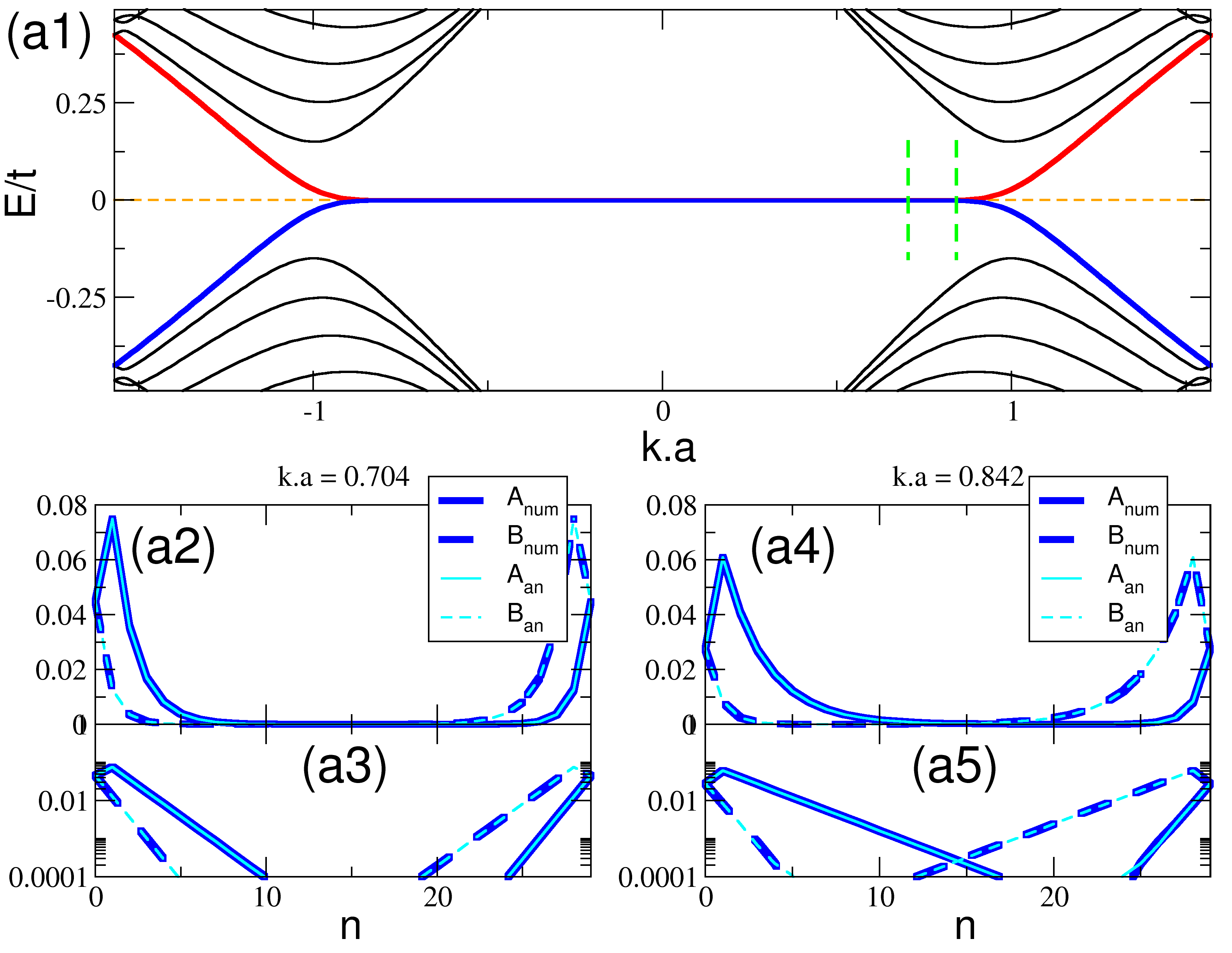}

\includegraphics[width=0.49\textwidth]{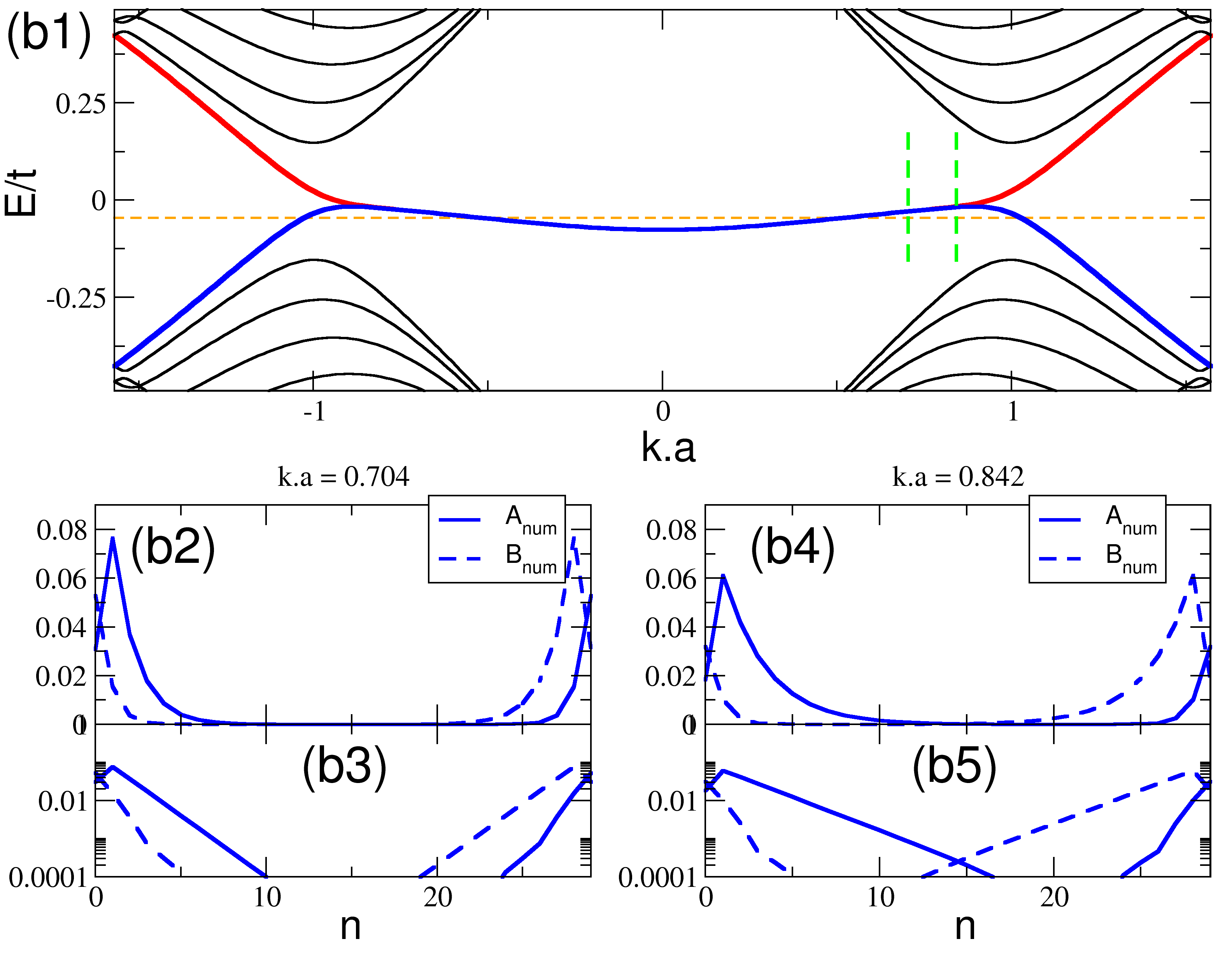}
\caption{$(a)$: ZGNR with two simplified $zz(57)$ edges ($h_{i}=v=h_{i}'=1$)
and a width of $65\mathring{A}$ (or $N=30$ zigzag lines). The panel
$(a1)$ shows the tight-binding low-energy spectrum in the FBZ. The
two lowest-energy levels are colored in blue and red. The dashed (orange)
horizontal line, signals the position of the Fermi level. Panels $(a2)$
and $(a4)$, show the edge state squared amplitude corresponding to
the blue level in $(a1)$, for $ka=0.704$ and $ka=0.842$ respectively
(whose position is identified in panel $(a)$ by the vertical dashed
green lines). The continuous (dashed) dark blue line stands for the
amplitude in the $A_{1}$-sub-lattice ($B_{1}$-sub-lattice) corresponding
to the blue level in $(a1)$ obtained from the numerical diagonalization
of the TB Hamiltonian (the red level is an identical edge state).
Only the amplitudes $A_{1}(k;n)$ and $B_{1}(k;n)$ were plotted,
because $A_{2}(k;n)$ and $B_{2}(k;n)$ are identical to the former.
The continuous (dashed) light blue line stands for the zero-energy
edge state amplitude in the $A_{1}$-sub-lattice ($B_{1}$-sub-lattice)
obtained analytically in a semi-infinite ribbon. Note the extreme
coincidence between the numerical and the analytical edge states.
Panels $(a3)$ and $(a5)$, show the same plots as in $(a2)$ and
$(a4)$, but now with logarithmic scale in the $y$-axis, to display
the exponential decay of the squared amplitudes. $(b)$: ZGNR with
two real $zz(57)$ edges (see Table \ref{tab_hopp}) and a width of
$65\mathring{A}$. The panels are organized as those of $(a)$.}

\label{fig:TB_SWD-57_Spectrum-EdgeStates_WithAndWithoutCorrections} 
\end{figure}

In Fig. \ref{fig:TB_SWD-57_Spectrum-EdgeStates_WithAndWithoutCorrections}
we compare numerical diagonalization results with those of the present
analysis, for the simplified situation where hopping renormalizations
at the edge are ignored, $h_{i}=v=h_{i}'=1$.%
\footnote{We have also confirmed numerically the prediction that edges states
have zero energy when $h_{1}^{2}-h_{2}h_{4}=0$, though we do not
present the corresponding data. %
} The squared amplitudes of the edge states, of a narrow ribbon with
$N=30$ ($65\mathring{A}$ wide), calculated numerically, are indeed
in very good agreement with those of the edge states of a semi-infinite
ribbon obtained analytically, from Eqs. (\ref{eq:wfunctions_zero_energy})
and (\ref{eq:wfunctions_zero_energy-c}). 

Unlike the zero energy states occurring in unreconstructed ZGNR, the
wave function amplitudes of the edge states are non-zero in both sub-lattices.
Those familiar with the Dirac equation description of graphene might
find this result surprising, since, at zero energy, the equations
for the $A$ and $B$ fields decouple, and only one of them can be
non-zero.\cite{BF06} However, as can be seen in Fig. \ref{fig:TB_SWD-57_Spectrum-EdgeStates_WithAndWithoutCorrections},
panels $(a4)$-$(a5)$ -- which refer to a value of $k$ close to
a Dirac point --, the decay length is much shorter in the $B$ sub-lattice;
this is related to the fact that the $B$ amplitudes correspond to
the $\sigma=+$ mode, which, in the bulk, is high energy, and has
a \emph{finite} decay length, of the order of a single row width,
\emph{even} at the Dirac point, contrasting with the $\sigma=-$ mode,
whose decay length diverges at the Dirac point. So, away from the
boundary, the edge state wave function is, in fact, similar to that
of a ZGNR, because the amplitude at the $B$ sub-lattice is exponentially
smaller than in the $A$ one; but the reconstructed edge requires
the presence of the confined $\sigma=+$ mode, in order to satisfy
the BC. When we move away from the Dirac point, the distinction between
high and low energy modes washes away, and both modes are confined
within atomic distances to the edges {[}Fig. \ref{fig:TB_SWD-57_Spectrum-EdgeStates_WithAndWithoutCorrections},
panels $(a2)$-$(a3)${]}.

At this point we come back to the consideration of real edges, where
the hopping parameters have the values in Table~\ref{tab_hopp}.
One does not find $\mathcal{M}_{++}=0,$ and the BCs of Eqs. (\ref{eq:bc_main_text})
are no longer compatible with the conditions for zero energy states,
$\alpha_{+}(k)=\beta_{-}(k)=0$, (or $\alpha_{+}(k)=\alpha_{-}(k)=0$,
if $\left|\xi_{A}^{-}\right|>1$); edge states, if they exist, have
to be dispersive. The dispersiveness of the edge states' levels can
be seen in panel $(b1)$ of Fig. \ref{fig:TB_SWD-57_Spectrum-EdgeStates_WithAndWithoutCorrections}.

We analyze this situation in detail in Appendix \ref{sec:LowEnEdgeStates}.
If a semi-infinite 1D $AB$ chain has a zero energy edge state with
BC, say $B(0)=0$, it will still have a low but finite energy one,
if the BC is replaced by $B(0)=sA(0)$ with $\left|s\right|<1.$ In
the present case the situation is more complex, because the problem
involves two 1D chains {[}Eqs. (\ref{eq:tB_diagonal_chain}){]}, coupled
by the BC {[}Eq. (\ref{eq:bc_main_text}){]}. The main conclusion
still holds, and we expect low energy, dispersive, edge states near
the Dirac points ($ka=\pm\pi/3$).
\begin{figure}
\includegraphics[width=1\columnwidth]{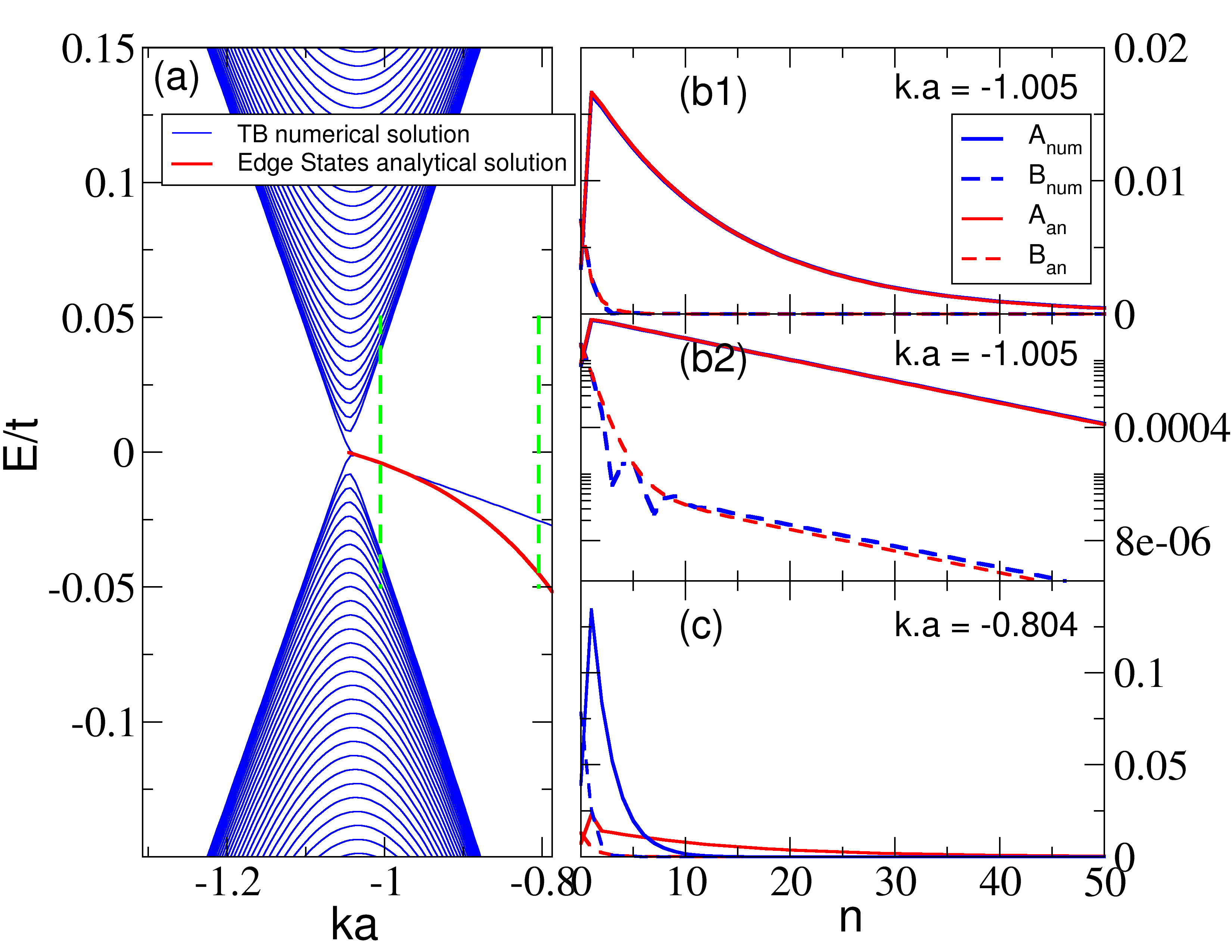}
\caption{Comparison between the edge state levels obtained from numerical diagonalizing
the tight-binding Hamiltonian of a ribbon with $N=600$ ($\approx900\mathring{A}$
wide) (in blue), and the edge state level resulting from analytically
solving the TB equations, by using simplified (zero-energy) BCs (in
red). $(a)$ shows the energy as function of $k$; panels $(b1)$
and $(c)$, show the square of the amplitudes as function of distance
to the edge, for $ka=-1.005$ and $ka=-0.804$ respectively. These
values of $ka$ are identified in panel $(a)$ by vertical green dashed
lines; $(b2)$ panel is the same as $(b1)$, but now in logarithmic
scale.}
\label{fig:Comp_TB_Analytical_NonZeroES-1}
\end{figure}

In Fig. \ref{fig:Comp_TB_Analytical_NonZeroES-1} we compare analytical
results for a semi-infinite chain, obtained with the procedure described
in Appendix \ref{sec:LowEnEdgeStates}, with numerical diagonalization
of a very wide ribbon ($N=600$). The use of the zero energy BC of
Eq. (\ref{eq:bc_main_text}) correctly accounts for the wave function
and for the energy dispersion as a function of $k$, but only very
close to the Dirac point. It quickly deviates strongly from the numerical
results as we move away from the Dirac point. This is to be expected,
not only as a result of the violation of the low energy condition,
but, more importantly, because the description in terms of bulk equations
and simplified BCs will not hold when the edge state has such a short
decay length that it lives mostly at the edge. Moreover, as stated
before, near the Dirac points the localization length of the mode
$\sigma=-$ diverges. As a consequence, the analytical analysis developed
in Appendix \ref{sec:LowEnEdgeStates}, will only accurately describe
the physics of $zz(57)$ edged ribbons near the Dirac points if the
ribbons are large.

We can summarize the results of this subsection, saying that, as a
consequence of the duplication of the unit cell, Stone-Wales reconstructed
edges present a new type of edge state, 
\begin{align}
\mathbf{A}(k;n) & =e^{iq_{+}(n-2)}\alpha_{+}\mathbf{u}_{+}+\alpha_{-}e^{iq_{-}(n-2)}\mathbf{u}_{-},\\
\mathbf{B}(k;n) & =e^{iq_{+}(n-2)}\beta_{+}\mathbf{u}_{+}+\beta_{+}e^{iq_{-}(n-2)}\mathbf{u}_{-},
\end{align}
with the following features: (i) the states are dispersive; (ii) the
wave-function, even for the semi-infinite ribbon, has non-zero amplitudes
on both sub-lattices; (iii) close to $k=\pm\pi/3$, the Dirac points,
the wave function amplitudes have two components decaying with very
different rates, $\Im q_{-}$ and $\Im q_{+}$, the latter remaining
finite even at the Dirac point, and corresponding to a mode with only
atomic scale penetration into the bulk. 

This last feature is strikingly apparent in Fig. \ref{fig:Comp_TB_Analytical_NonZeroES-1},
panel $(b2)$, where the faster decaying component in the $B$ lattice
dominates the wave function close to the edge, because of a larger
initial amplitude, $\left|\beta_{+}\right|\gg\left|\beta_{-}\right|$,
but is supplanted by the one with slower decay, around $n\approx10$.


\subsection{Perpendicular magnetic field}

\label{subsec:TB_landau}

When a perpendicular magnetic field is applied to a graphene sheet,
electrons acquire a cyclotron motion, with quantized cyclotron radius
and macroscopically degenerate energy levels, the so called Landau
levels (LL). In a ribbon, LL degeneracy is partially lifted, because
the edges interrupt the cyclotron orbits located close to them. In
this section we discuss the effect of a perpendicular magnetic field
in the low energy spectrum of the tight-binding models we have been
discussing (a ribbon with a $zz(57)$ reconstruction).
\begin{figure}[htp!]
\includegraphics[width=1\columnwidth]{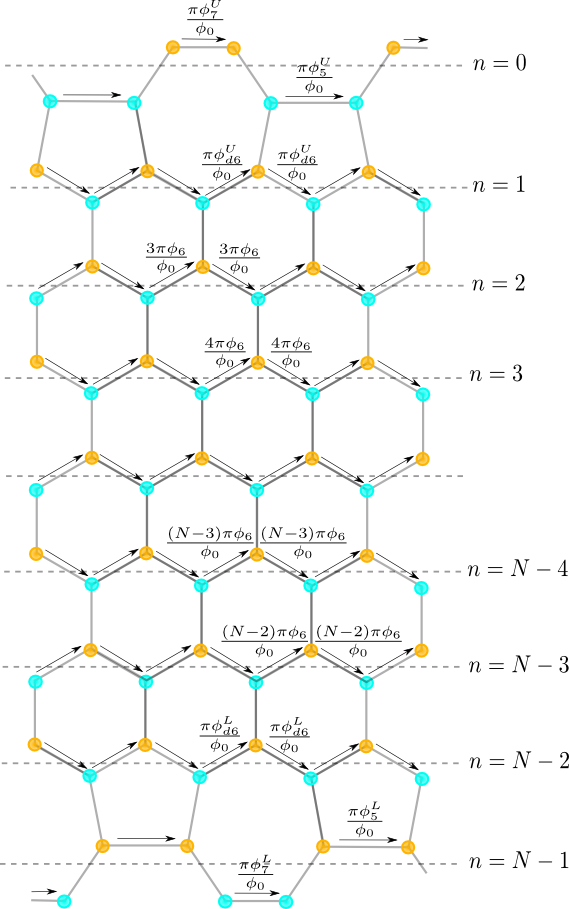}
 \caption{Peierls phases of a zigzag ribbon with totally reconstructed edges
($N=10$).}

\label{fig:PeierlsPhases} 
\end{figure}

The introduction of a static magnetic field, applied perpendicularly
to the ribbon, $\boldsymbol{B}=B\boldsymbol{\hat{e}_{z}}$, can be
achieved by a Peierls substitution of the hopping integrals,\cite{Peierls_ZPhys:1933,Boykin_PRB:2001}
\begin{eqnarray}
t_{ij} & \to & t_{ij}e^{i2\pi\phi_{ij}},\label{eq:PeierlsSubstitution}
\end{eqnarray}
 where $t_{ij}$ stands for the hopping integral between the position
$\boldsymbol{R_{i}}$ and the position $\boldsymbol{R_{j}}$ in the
absence of a magnetic field, and the phase $\phi_{ij}$ is given by
the line integral 
\begin{eqnarray}
\phi_{ij} & = & \frac{1}{\phi_{0}}\int_{\boldsymbol{R}_{i}}^{\boldsymbol{R}_{j}}\boldsymbol{A}\cdot\textrm{d}\boldsymbol{r},\label{eq:PeierlsPhase}
\end{eqnarray}
where $\boldsymbol{A}$ is the potential vector and $\phi_{0}=h/e$
is the flux quantum. Note that the magnetic flux through the area
$\Sigma$, in units of the flux quantum $\phi_{0}$, is 
\begin{eqnarray}
\frac{1}{\phi_{0}}\int_{\Sigma}\textrm{d}\boldsymbol{\sigma}\cdot\boldsymbol{B} & = & \frac{1}{\phi_{0}}\oint\textrm{d}\boldsymbol{r}\cdot\boldsymbol{A}=\sum_{\mathrm{around}\textrm{ }\Sigma}\phi_{ij}.\label{eq:ClosurePeierlsPhase}
\end{eqnarray}

The zigzag edge reconstruction modifies, not only the hoppings, but
also the areas of the pentagons, heptagons and hexagons near the edge.
Therefore, by Eq. (\ref{eq:PeierlsPhase}), the Peierls phases around
the edges are distinct from those in the ribbon bulk. We choose a
gauge that yields Peierls' phases as shown in Fig. \ref{fig:PeierlsPhases},
clearly satisfying Eq. (\ref{eq:ClosurePeierlsPhase}), $\phi_{6}$
being the magnetic flux per hexagon in the bulk graphene lattice.

\begin{figure}[htp!]
\includegraphics[width=0.99\columnwidth]{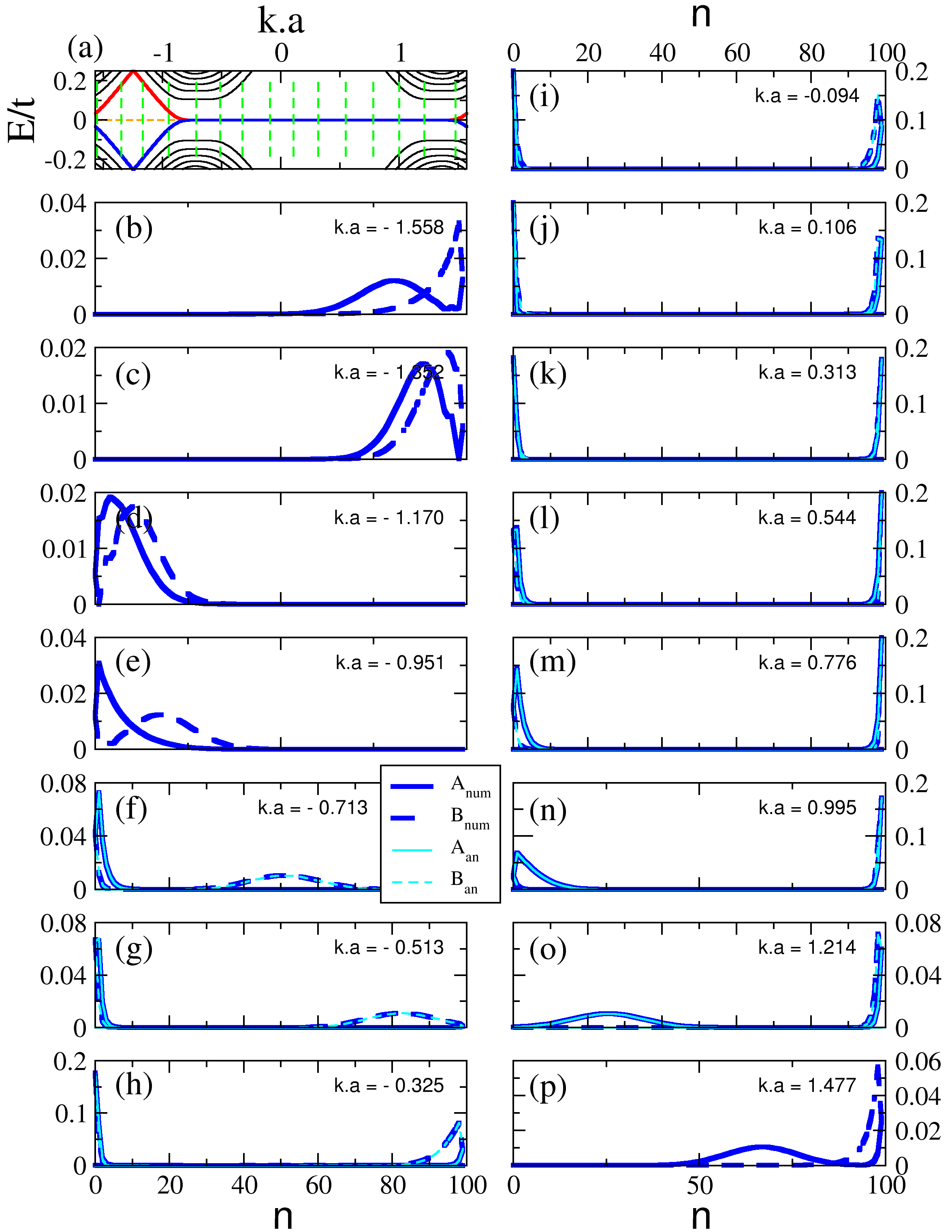}
\caption{Panel $(a)$ shows the tight-binding energy spectrum of a zigzag graphene
nanoribbon with simplified $zz(57)$ edges ($h_{i}=v=h'_{i}=1$ and
$\phi_{5}^{C}=\phi_{7}^{C}=2\phi_{\textrm{d}6}^{C}=2\phi_{6}$) with
a width of $214\mathring{A}$ (or $N=100$ zigzag rows) in the presence
of a perpendicular magnetic field, $B=80T$. The green dashed vertical
lines in $(a)$, indicate the different values of $ka$ for which
the edge states were plotted in $(b)$-$(p)$. Panels $(b)$-$(p)$
show in dark blue, for different values of $ka$, the wave function
squared modulus of the two lowest-energy levels highlighted in panel
$(a)$ with blue and red fill. The light blue curves in panels $(f)$-$(o)$
stand for the edge states obtained analytically for values of $ka$
for which their energy is zero {[}see panel $(a)${]}.}

\label{fig:LLwithoutHopps} 
\end{figure}

The spectrum shown in Fig.~\ref{fig:LLwithoutHopps}$(a)$ is essentially
the same as for a pristine ZGNR (apart from the folding of the Brillouin
zone), the most prominent feature being a doubly degenerate zero energy
level occurring between the two Dirac points. But what is displayed
is, in fact, the spectrum of a ribbon with simplified $zz(57)$ edges,
where hopping renormalizations were ignored ($h_{i}=1$, $v_{i}=1$),
and the pentagons and heptagons considered to have the same area as
all the hexagons.

\begin{figure}[htp!]
\includegraphics[width=0.99\columnwidth]{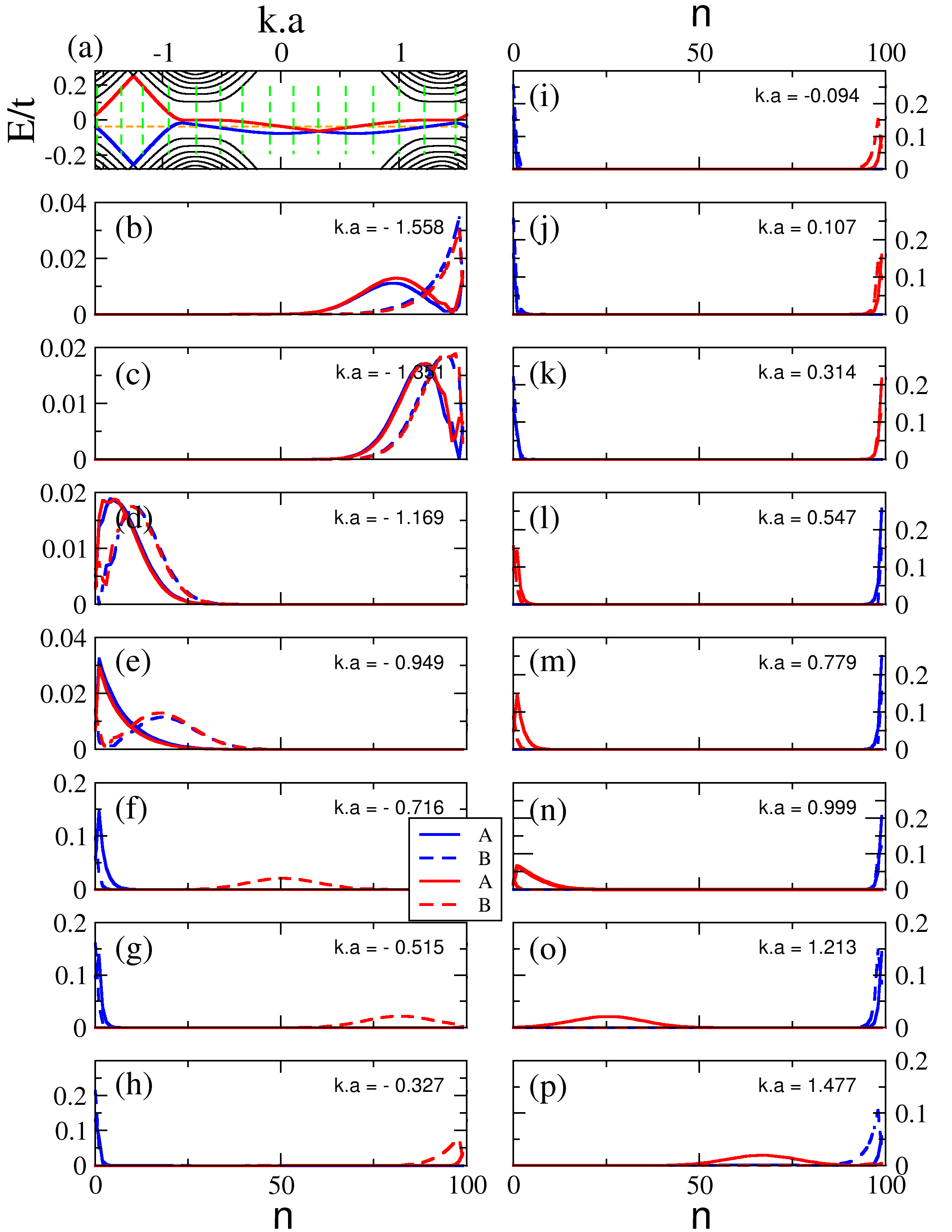}
\caption{Panel $(a)$ shows the tight-binding energy spectrum of a zigzag graphene
nanoribbon with real $zz(57)$ edges with a width of $214\mathring{A}$
(or $N=100$ zigzag rows) in the presence of a perpendicular magnetic
field, $B=80T$. Panels $(b)$-$(p)$ show the squared modulus of
the lowest-energy levels wave functions for different values of $ka$
{[}blue and red levels in panel $(a)${]}. The green dashed vertical
lines in $(a)$, indicate the different values of $ka$ for which
the edge states wave functions squared modulus were plotted in $(b)$-$(p)$.}

\label{fig:LLwithHopps} 
\end{figure}

In Fig. \ref{fig:LLwithHopps}$(a)$, we display the spectrum of a
zigzag ribbon with real $zz(57)$ edges in the presence of a perpendicular
magnetic field. In contrast with the previous case, the two zero-energy
levels are now split in energy and dispersive, crossing each other
at the $\Gamma$-point.%
\footnote{The whole spectrum is shifted in $ka$, because we have set $n=0$
at the upper edge (where $n$ is the label of the zigzag rows). If
we have set $n=0$ to the center of the ribbon, the shift would disappear.\cite{Wakabayashi_PRB:1999}%
} 

The plots of the wave functions suggest a clear interpretation of
this result. In graphene there is a bulk zero energy LL which cannot
be affected by BCs, because the corresponding wave functions are localized
in the bulk and do not reach the edges. And, in fact, one can see
in Fig.~\ref{fig:LLwithHopps}$(a)$ regions of $k$ with a flat
energy level at zero energy; the plots of the corresponding wave functions
($-0.9\lesssim ka\lesssim-0.5$ and $1.2\lesssim ka\lesssim1.5$)
show one localized state inside the ribbon. The remaining states are
edge states localized at its boundaries. In a real reconstructed edge,
these states are dispersive in zero magnetic field, as we have seen
in the previous section, and remain dispersive in a magnetic field:
hence the lifting of the degeneracy and the level crossing at the
$\Gamma$ point, which involves states localized at opposite edges.
On the other hand, in the simplified $zz(57)$ ribbon, the edge states
occur at zero energy, as we have also seen. So the doubly degenerate
zero energy state is either a zero energy bulk LL and an edge state,
or \emph{two} edge states, located at opposite ends of the ribbon.
This is confirmed by the plots of the wave functions.

\begin{figure}[htp!]
\includegraphics[width=0.99\columnwidth]{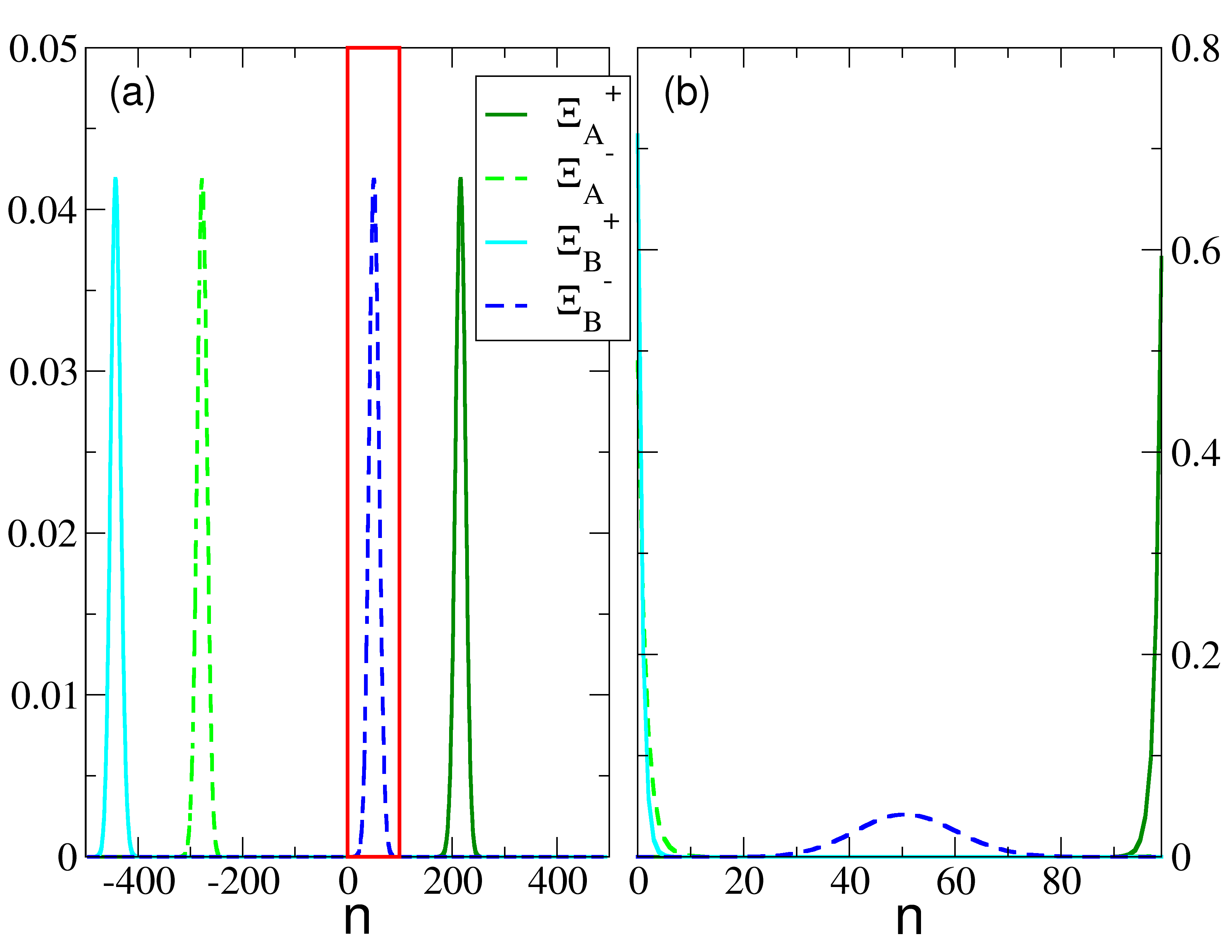}
\caption{Plot of $|\Xi_{A}^{\pm}(n)|$ and $|\Xi_{B}^{\pm}(n)|$ (that can
be interpreted as the amplitudes of the four components of the wave
function in the proper basis of the matrices $\widetilde{W}$), given
by Eqs. (\ref{eq:Xis}). The above quantities were plotted for $B=80T$
and $ka=-0.716$. In the $(a)$ panel, the $|\Xi_{A/B}^{\pm}(n)|$
are normalized over the region $n\in[-500,500]$, while in the $(b)$
panel they are normalized over $n\in[0,100]$. The red box in panel
$(a)$ signals the region where $n\in[0,100]$.}

\label{fig:MagFldRecRelGeneral} 
\end{figure}

We now proceed to indicate briefly how these results arise from the
Peierls substitution. We begin by considering the appearance of a
zero energy bulk Landau level (BLL). The recurrence relations for
the amplitudes now involve matrices that depend on the row index,
\begin{subequations}\label{eq:MagRecurrence-1} 
\begin{eqnarray}
\boldsymbol{A}(k;n+1) & = & \widetilde{W}_{A}(n)\boldsymbol{A}(k;n),\label{eq:MagRecurrence_bulk1-1}\\
\boldsymbol{B}(k;n+1) & = & \widetilde{W}_{B}^{-1}(n+1)\boldsymbol{B}(k;n).\label{eq:MagRecurrence_bulk2-1}
\end{eqnarray}
 \end{subequations} Recall that $\boldsymbol{A}(k;n)$ and $\boldsymbol{B}(k;n)$
are notations for the column vectors, $\boldsymbol{A}(k;n)=\big[A_{1}(k;n),A_{2}(k;n)\big]^{T}$
and $\boldsymbol{B}(k;n)=\big[B_{1}(k;n),B_{2}(k;n)\big]^{T}$; the
matrices $\widetilde{W}_{A}(n)$ and $\widetilde{W}_{B}(n)$ are written
in Appendix \ref{sec:MatricesAppndxB1}. As before, these are commuting
matrices, and have the common basis $\{\mathbf{u_{+}},$$\mathbf{u}_{-}\}$;
the eigenvalues, however, depend on the row index, \begin{subequations}\label{eq:xiGen}
\begin{eqnarray}
\widetilde{\xi_{A}^{+}}(r) & = & -2e^{ika/2}\cos\left[\frac{ka}{2}-(r+1)\pi\frac{\phi_{6}}{\phi_{0}}\right]=\left(\widetilde{\xi_{B}^{+}}(r)\right)^{*},\nonumber \\
\\
\widetilde{\xi_{A}^{-}}(r) & = & 2ie^{ika/2}\sin\left[\frac{ka}{2}-(r+1)\pi\frac{\phi_{6}}{\phi_{0}}\right]=\left(\widetilde{\xi_{B}^{-}}(r)\right)^{*}.\nonumber \\
\end{eqnarray}
 \end{subequations}We can then rewrite Eqs. (\ref{eq:MagRecurrence-1}),
for $m\geq2,$ as \begin{subequations} \label{eq:MagRecurrProp}
\begin{eqnarray}
\boldsymbol{A}(k;n) & = & \Xi_{A}^{+}(n,m)\alpha_{+}(k;m)\boldsymbol{u^{+}}\nonumber \\
 & + & \Xi_{A}^{-}(n,m)\alpha_{-}(k;m)\boldsymbol{u^{-}},\label{eq:MagArecurrProp}\\
\boldsymbol{B}(k;n) & = & \Xi_{B}^{+}(n,m)\beta_{+}(k;m)\boldsymbol{u^{+}}\nonumber \\
 & + & \Xi_{B}^{-}(n,m)\beta_{-}(k;m)\boldsymbol{u^{-}},\label{eq:MagBrecurrProp}
\end{eqnarray}
 \end{subequations} where $n\geq m$, $\alpha_{\sigma}$ and $\beta_{\sigma}$
are undetermined coefficients, while the quantities $\Xi_{A}^{\pm}(n,m)$
and $\Xi_{B}^{\pm}(n,m)$ are a shorthand for \begin{subequations}
\label{eq:Xis} 
\begin{eqnarray}
\Xi_{A}^{\sigma}(n,m) & = & \prod_{r=m}^{n-1}\widetilde{\xi_{A}^{\sigma}}(r),\\
\Xi_{B}^{\sigma}(n,m) & = & \prod_{r=m+1}^{n}\frac{1}{\widetilde{\xi_{B}^{\sigma}}(r)}.
\end{eqnarray}
 \end{subequations}As a function of the row index $n$, $\Xi_{A(B)}^{\sigma}(n,m)$
goes through a maximum when $\left|\widetilde{\xi_{A}^{\sigma}}(r)\right|\left(\left|\widetilde{\xi_{B}^{\sigma}}(r)\right|^{-1}\right)$
decreases below 1. These maxima are repeated periodically when $n$
changes by $2n_{\phi}$, where $n_{\phi}\equiv\phi_{0}/\phi_{6}$
is the number of hexagons required for a total flux equal to a flux
quantum. These multiple maxima are related to commensurability effects
between the lattice parameter and the cyclotron radius and are only
important for unrealistically high fields. \cite{Thouless_PRL:1982}
For achievable values of the magnetic field, $n_{\phi}$ is much larger
than the ribbon width, $N$, (for $B=80\,\mathsf{T}$, $2n_{\phi}\approx2000$),
and at most one maximum of $\Xi_{A(B)}^{\sigma}(n,m)$ is located
inside the ribbon, as shown in Fig. \ref{fig:MagFldRecRelGeneral}.
Assume, for instance, that that is the case for $\Xi_{B}^{-}$ at\begin{subequations}\label{eq:MaxXis}

\begin{align}
\bar{n}_{B-} & =\frac{ka}{2\pi}n_{\phi}-\left(\frac{5}{6}+q\right)n_{\phi},\qquad1\ll\bar{n}_{B-}\ll N-1,\label{eq:MaxXiBm}\\
\bar{n}_{A+} & =\frac{ka}{2\pi}n_{\phi}-\left(\frac{2}{3}+q\right)n_{\phi}=\bar{n}_{B-}+\frac{n_{\phi}}{6},\label{eq:MaxXiAp}\\
\bar{n}_{A-} & =\frac{ka}{2\pi}n_{\phi}-\left(\frac{1}{6}+q\right)n_{\phi}=\bar{n}_{B-}+\frac{2n_{\phi}}{3},\label{eq:MaxXiAm}\\
\bar{n}_{B+} & =\frac{ka}{2\pi}n_{\phi}-\left(\frac{1}{3}+q\right)n_{\phi}=\bar{n}_{B-}+\frac{n_{\phi}}{2},\label{eq:MaxXiBp}
\end{align}
\end{subequations}where $q$ is an integer. From Eqs. (\ref{eq:MaxXis}),
we conclude that for reasonable values of the magnetic field and ribbon
widths, at most, only one of the components will have a maximum inside
the ribbon (of width $N=100$). See, as an example, Fig. \ref{fig:MagFldRecRelGeneral}.

Moreover, the amplitude $\beta_{-}(k,n)$ will decay exponentially
to very small values at the edges; to exponential accuracy, the BCs,
whatever they may be, are trivially satisfied by choosing $\alpha_{+}=\alpha_{-}=\beta_{+}=0$;
this then is a BLL, where the wave function exists only in one of
the sub-lattices and is localized away from the edges. These BLLs
occur irrespective of the type of edge. However, when $k$ changes
and the LL center approaches the edge, the BCs come into play, differentiating
the various situations.

Let us now consider the appearance of the edge states in these results.
The general BC for a reconstructed zigzag edge with SW defects at
the $n=0$ end, may be written as $\boldsymbol{\alpha}(k;2)=\widetilde{\mathcal{M}}\mathbf{\boldsymbol{\beta}}(k;2)$,
where $\widetilde{\mathcal{M}}$ is defined in Appendix \ref{sec:MatricesAppndxB1},
Eq. (\ref{eq:BC_MagF}), with an analogous expression for the edge
in $n=N-1$, $\mathbf{\boldsymbol{\beta}}(k;N-3)=\widetilde{\mathcal{M}}'\boldsymbol{\alpha}(k;N-3)$.

We will start by assuming that the ribbon is terminated with a simplified
$zz(57)$ edge ($h_{i}=v=h_{i}'=1$ and $\phi_{7}=\phi_{5}=2\phi_{d6}=2\phi_{6}$).
In such a situation, we have $\widetilde{\mathcal{M}}_{++}=\widetilde{\mathcal{M}}_{--}=0$,
a result which uncouples the components $\alpha_{+}$ and $\beta_{-}$
from $\alpha_{-}$ and $\beta_{+}$ \begin{subequations}\label{eq:MagBCuncoup}
\begin{eqnarray}
\alpha_{+}(k;2) & = & \widetilde{\mathcal{M}}_{+-}\beta_{-}(k;2),\\
\alpha_{-}(k;2) & = & \widetilde{\mathcal{M}}_{-+}\beta_{+}(k;2).
\end{eqnarray}
\end{subequations}

As a consequence, every time we have a zero energy BLL (living away
from the edges), we will also have one other solution of zero energy,
now localized at the edge. Let us take as an example, the case where
$ka=-0.716$, for which the $\Xi_{A/B}^{\pm}$ are depicted in Fig.
(\ref{fig:MagFldRecRelGeneral}). To exponential accuracy, the BCs
involving $\beta_{-}$ and $\alpha_{+}$ are trivially satisfied at
both edges choosing $\alpha_{+}=0$. Those involving $\beta_{+}$
and $\alpha_{-}$, are satisfied at the upper edge choosing $\alpha_{-}(k;2)=\widetilde{\mathcal{M}}_{-+}\beta_{+}(k;2)$,
being satisfied at the lower edge to exponential accuracy. In the
real space {[}see Eqs. (\ref{eq:MagRecurrProp}){]}, we will have
a BLL localized only on the $B$ sub-lattice and an edge state around
the edge at $n=0$, living in both sub-lattices with different localization
lengths.

When the value of $ka$ is increased, the maxima of $\Xi_{A/B}^{\pm}$
move to higher values of $n$. At a certain point, the maximum of
$\Xi_{B}^{-}$ will be such that $\bar{n}_{B-}>N-1$, and then there
will be no maxima inside the ribbon. In such a case, the maxima of
$\Xi_{B}^{-}$ and $\Xi_{A}^{+}$ closer to the ribbon, will be at
$n>N-1$, while the maxima of $\Xi_{B}^{+}$ and $\Xi_{A}^{-}$ closer
to the ribbon, will be at $n<0$. In this case, the BCs involving
$\beta_{-}$ and $\alpha_{+}$ will be satisfied at the lower edge
choosing $\beta_{-}(k;N-3)=\widetilde{\mathcal{M}}'_{-+}\alpha_{+}(k;N-3)$.
At the upper edge, the BC will be obeyed to exponential accuracy.
The converse needs to be done regarding the BCs involving $\beta_{+}$
and $\alpha_{-}$. Consequently, for $-0.5\lesssim ka\lesssim1.2$,
there will be zero-energy solutions localized at both edges, living
in both sub-lattices with distinct localization lengths in each sub-lattice.

If, on the contrary, we start decreasing the value of $ka$ from $ka\approx-0.72$,
the maxima of $\Xi_{A/B}^{\pm}$ moves to lower values of $n$, and
at a certain point, the maximum of $\Xi_{B}^{-}$ will be such that
$\bar{n}_{B-}<0$. In such case, the maxima of $\Xi_{B}^{-}$ and
$\Xi_{A}^{-}$ closer to the ribbon, will be at $n<0$, while the
maxima of $\Xi_{B}^{+}$ and $\Xi_{A}^{+}$ closer to the ribbon,
will be at $n>N-1$. In this case, it will not be possible to satisfy
the BCs non-trivially and consequently, there will be  no zero-energy
solutions in this region, as can be seen in Fig. \ref{fig:LLwithoutHopps}$(a)$. 

When instead of simplified $zz(57)$ edges, the ribbon is terminated
with real $zz(57)$ edges, the matrix $\widetilde{\mathcal{M}}$ is
modified, and $\widetilde{\mathcal{M}}_{++}\neq\widetilde{\mathcal{M}}_{--}\neq0$,
resulting in a BC coupling all the components $\alpha_{\pm}$ and
$\beta_{\pm}$ \begin{subequations}\label{eq:MagBCcoup}

\begin{eqnarray}
\alpha_{+}(k;2) & = & \widetilde{\mathcal{M}}_{++}\beta_{+}(k;2)+\widetilde{\mathcal{M}}_{+-}\beta_{-}(k;2),\\
\alpha_{-}(k;2) & = & \widetilde{\mathcal{M}}_{-+}\beta_{+}(k;2)+\widetilde{\mathcal{M}}_{--}\beta_{-}(k;2).
\end{eqnarray}
\end{subequations} 

To grasp the implications of this modification, consider for instance
the case where $ka=-0.716$, depicted in Fig. \ref{fig:MagFldRecRelGeneral},
where a BLL is present in the $\beta_{-}$ mode; since $\beta_{-}(2)\approx0,$
the BC imply all three remaining amplitudes, $\alpha_{+},\,\alpha_{-}$and
$\beta_{+}$ to be non zero, if there is to be an edge state in addition
to the BLL. But the $\alpha_{+}$ mode grows as $n$ increases, whereas
the other two decrease; as a result the BCs will be violated at the
opposing edge. In conclusion, BCs can no longer be satisfied with
zero energy edge states, which become dispersive, whereas zero energy
BLL still occur. This accounts for the lack of zero energy doubly
degenerate state in ribbons with real reconstructed $zz(57)$ edges.


\section{Conclusion}

We have discussed in detail the effect of edge reconstruction on the
characteristics of low energy edge states in graphene ribbons. In
the case of Stone-Wales $zz(57)$ reconstructed zigzag edges, we find
a new type of edge state originating from the doubling of the unit
cell along the edge, brought about by the edge reconstruction. This
new type of edge state has the following features: (i) the states
are in general dispersive, although specific values of the tight-binding
model parameters allow zero energy states; (ii) the wave-function,
even for the semi-infinite ribbon, has non-zero amplitudes on both
sub-lattices; (iii) close to the Dirac points, the wave function amplitudes
have two components decreasing with the distance from edge with different
decay lengths, one of which remains finite, of the order of the lattice
parameter, even at the Dirac point, while the other diverges. The
dispersion of the edge states should lead to a charge transfer between
bulk and edges (self-doping), which, for realistic values of the tight-binding
parameters, leaves the edges negatively charged.

In the presence of a magnetic field, one still finds zero energy bulk
Landau Levels, as was to be expected, since these are insensitive
to the edges; however, in contrast to pristine zigzag ribbons, where
the zero energy LL is degenerate with an edge state, this in no longer
true in ribbons with reconstructed edges, since the edge states are,
in general, dispersive.


\begin{acknowledgements}

J. N. B. R. was supported by Fundação para a Ciência e a Tecnologia
(FCT) through Grant No. SFRH/BD/44456/2008. N. M. R. P. and R. M.
R. were supported by Fundos FEDER through the Programa Operacional
Factores de Competitividade - COMPETE and by FCT under project no.
PEst-C/FIS/UI0607/2011.

\end{acknowledgements}

\appendix


\section{Tight Binding equations and boundary conditions\label{sec:appendix}}

In this appendix, we write the tight-binding equations for the amplitudes
at the sites near one edge, $n=0$; these will determine the boundary
conditions (BCs) that must be satisfied by the bulk solutions. For
clarity, we begin by considering zero energy states. We will argue
that the BCs adequate for low energy states, $\left|\epsilon/t\right|\ll1$,\emph{
are the same as for zero energy states}.

The tight-binding equations at the sites of $A_{1}(m,0)$ and $A_{2}(m,0)$
have the form:\begin{subequations}

\begin{eqnarray}
h_{2}A_{2}(m;0)+h_{1}B_{1}(m;0) & = & 0,\label{eq:tB_A0}\\
h_{2}A_{1}(m,0)+h_{1}B_{2}(m;0) & = & 0.\label{eq:TB_A0_1}
\end{eqnarray}
\end{subequations} It will be useful to express these in matrix form;
after Fourier transforming in the $m$ index, ($k$ is the wave vector
along the edge),

\begin{equation}
\mathbf{A}(k;0)=-\frac{h_{1}}{h_{2}}\sigma_{x}\mathbf{B}(k;0),\label{eq:tB_A0_matrix}
\end{equation}
 where $\sigma_{x}$ is a Pauli matrix. For the $B_{1}(m;0),\, B_{2}(m;0)$
sites, \begin{subequations}\label{eq:tB_B0}
\begin{align}
h_{4}B_{2}(m-1;0)+h_{1}A_{1}(m;0)+vA_{1}(m;1) & =0,\label{eq:tB_B0_1-1}\\
h_{4}B_{1}(m+1;0)+h_{1}A_{2}(m;0)+vA_{2}(m;1) & =0.\label{eq:tB_B0_2-1}
\end{align}
\end{subequations}Using Bloch's theorem, we can cast this in the
form 
\begin{align}
\mathbf{A}(k;1) & +\frac{h_{1}}{v}\mathbf{A}(k;0)\nonumber \\
 & +\frac{h_{4}}{v}\left[\begin{array}{cc}
e^{-2ika} & 0\\
0 & e^{2ika}
\end{array}\right]\sigma_{x}\mathbf{B}(k;0)=0.
\end{align}
Using Eq. (\ref{eq:tB_A0_matrix}) in this one, 
\begin{equation}
\mathbf{A}(k;1)+\mathcal{R}\sigma_{x}\mathbf{B}(k;0)=0,\label{eq:tB_BO_matrix}
\end{equation}
where 
\begin{equation}
\mathcal{R}:=-\left[\begin{array}{cc}
\frac{h_{1}^{2}-h_{2}h_{4}e^{-2ika}}{h_{2}v} & 0\\
0 & \frac{h_{1}^{2}-h_{2}h_{4}e^{2ika}}{h_{2}v}
\end{array}\right],\label{eq:R_matrix}
\end{equation}
is a matrix that depends on $k$. 

With a similar procedure for the sites $A_{1}(m;1)$, $A_{2}(m;1)$,
$B_{1}(m;1)$ and $B_{2}(m,1)$, we obtain \begin{subequations}\label{eq:tB_row1}
\begin{eqnarray}
\mathbf{A}(k;2) & = & \mathcal{W}_{A}\mathbf{A}(k;1),\label{eq:tB_A1_matrix}\\
\mathbf{B}(k;0) & = & \mathcal{W}_{B}\mathbf{B}(k;1),\label{eq:tB_B1_matrix}
\end{eqnarray}
\end{subequations}with\begin{subequations}\label{eq:row1_matrices}
\begin{align}
\mathcal{W}_{A}= & -\left[\begin{array}{cc}
h'_{1} & h'_{1}\\
h'_{3}e^{2ika} & h'_{3}
\end{array}\right],\label{eq:WA_matrix}\\
\mathcal{W}_{B}= & -\frac{1}{v}\left[\begin{array}{cc}
h'_{1} & h'_{3}e^{-2ika}\\
h'_{1} & h'_{3}
\end{array}\right];\label{eq:WB_matrix}
\end{align}
\end{subequations} using Eqs. (\ref{eq:tB_BO_matrix}), we arrive
at
\begin{equation}
\mathbf{A}(k;2)+\mathcal{W}_{A}\mathcal{R}\sigma_{x}\mathcal{W}_{B}\mathbf{B}(k;1)=0.\label{eq:tB_BC1}
\end{equation}

Beyond the first row ($n>1$), it is simple to get\begin{subequations}\label{eq:recurrence_bulk_zero}
\begin{eqnarray}
\mathbf{A}(k;n+1) & = & W_{A}\mathbf{A}(k,n),\label{eq:bulk:recurrence_A}\\
\mathbf{B}(k;n+1) & = & W_{B}^{-1}\mathbf{B}(k,n),\label{eq:bulk_recurrence_B}
\end{eqnarray}
\end{subequations}where\begin{subequations}\label{eq:W_simple}
\begin{eqnarray}
W_{A} & = & -\left[\begin{array}{cc}
1 & 1\\
e^{2ika} & 1
\end{array}\right],\label{eq:Wa_simple}\\
W_{B} & = & -\left[\begin{array}{cc}
1 & e^{-2ika}\\
1 & 1
\end{array}\right],\label{eq:Wb_simple}
\end{eqnarray}
\end{subequations}

In summary, after Fourier transforming in the $m$ variable, the tight-binding
equations for a semi-infinite ribbon with $zz(57)$ reconstruction
are ($n>1$)\begin{subequations}\label{eq:tB_matrix_zero_energy}
\begin{eqnarray}
\mathbf{A}(k;2) & = & -\mathcal{W}_{A}\mathcal{R}\sigma_{x}\mathcal{W}_{B}W_{B}\mathbf{B}(k;2),\label{eq:tB_zero_energy_BC}\\
\mathbf{A}(k;n+1) & = & W_{A}\mathbf{A}(k,n),\label{eq:tB_A_recursion}\\
\mathbf{B}(k;n+1) & = & W_{B}^{-1}\mathbf{B}(k,n).\label{eq:tB_B_recursion}
\end{eqnarray}
\end{subequations} The last two are the bulk recursion relations,
while the first one contains the BC. 

We now generalize these equations for states of finite, but low, energy.
We argue that \emph{only the bulk equations are changed, the BCs remain
the same, i.e., }\begin{subequations}\label{eq:tB_matrix}
\begin{align}
\mathbf{A}(k;2) & =-\mathcal{W}_{A}\mathcal{R}\sigma_{x}\mathcal{W}_{B}W_{B}\mathbf{B}(k;2),\label{eq:tB_BC}\\
\mathbf{A}(k,n+1) & -W_{A}\mathbf{A}(k,n)=-\left(\frac{\epsilon}{t}\right)\mathbf{B}(k,n+1),\label{eq:tB_chain_A}\\
\mathbf{B}(k,n) & -W_{B}\mathbf{B}(k,n+1)=-\left(\frac{\epsilon}{t}\right)\mathbf{A}(k,n).\label{eq:tB_chain_B}
\end{align}
\end{subequations}Let us put back the energy in the equations for
the amplitudes near the edge, \begin{subequations}\label{eq:tB_A0_complete_Ant}
\begin{align}
h_{2}A_{2}(m;0)+h_{1}B_{1}(m;0) & =-\left(\frac{\epsilon}{t}\right)A_{1}(m;0),\label{eq:tB_A0_1_complete}\\
h_{2}A_{1}(m,0)+h_{1}B_{2}(m;0) & =-\left(\frac{\epsilon}{t}\right)A_{2}(m;0),\label{eq:tB_A0_2_complete}
\end{align}
 \end{subequations}so Eq.(\ref{eq:tB_A0_matrix}) becomes,

\begin{equation}
\mathbf{A}(m;0)+\frac{h_{1}}{h_{2}}\sigma_{x}\mathbf{B}(m;0)=\frac{1}{h_{2}}\left(-\frac{\epsilon}{t}\right)\sigma_{x}\mathbf{A}(m;0).\label{eq:tB_A0_complete}
\end{equation}
This shows the pattern that we have to repeat in Eqs. (\ref{eq:tB_B0})
through to Eqs. (\ref{eq:tB_row1}). Instead of Eq. (\ref{eq:tB_BC}),
we obtain, 
\begin{align}
\mathbf{A}(k;2)+ & \mathcal{W}_{A}\mathcal{R}\sigma_{x}\mathcal{W}_{B}W_{B}\mathbf{B}(k;2)\nonumber \\
=\left(-\frac{\epsilon}{t}\right) & \left[\frac{1}{v}\mathcal{W}_{A}\mathbf{B}(k;0)-\frac{h_{1}}{h_{2}v}\mathcal{W}_{A}\sigma_{x}\mathbf{A}(k;0)\right.\nonumber \\
- & \left.\frac{1}{v}\mathcal{W}_{A}\mathcal{R}\sigma_{x}\mathbf{A}(k;1)+\mathbf{B}(k,1)\right].\label{eq:tB_BC_complete}
\end{align}
Naturally, this reduces to Eq. (\ref{eq:tB_BC}) if the right hand
side is set to zero. The important point is that, for the values of
the parameters listed in Table \ref{tab_hopp}, the matrix $\mathcal{W}_{A}\mathcal{R}\sigma_{x}\mathcal{W}_{B}W_{B}$
has one finite eigenvalue in the entire range of $k$, whose modulus
is always larger than about 1.3. This means that, to lowest order
in $\left(-\epsilon/t\right)$, we are justified in neglecting the
RHS of this equation, and use the same BC as for zero energy states.
This is a valid approximation for states with $\left|\epsilon/t\right|\ll1$. 

Now we change basis to rewrite these equations in the eigenbasis of
$W_{A}$ and $W_{B}$, {[}see Eqs. (\ref{eq:preDiagRecurr-1}){]}\begin{subequations}\label{eq:u_pm}
\begin{eqnarray}
\mathbf{u}^{+} & = & \frac{1}{\sqrt{2}}\left[\begin{array}{c}
e^{-ika}\\
1
\end{array}\right],\label{eq:u_plus}\\
\mathbf{u}^{-} & = & \frac{1}{\sqrt{2}}\left[\begin{array}{c}
-e^{-ika}\\
1
\end{array}\right].\label{eq:u_minus}
\end{eqnarray}
\end{subequations} The coordinate transformation is defined by the
matrix $U$ given by 
\begin{equation}
U=\frac{1}{\sqrt{2}}\left[\begin{array}{cc}
e^{ika} & 1\\
-e^{ika} & 1
\end{array}\right].\label{eq:matrix_U}
\end{equation}
The BC in the new basis, becomes
\begin{align}
\boldsymbol{\alpha}(k;2) & =-U\mathcal{W}_{A}\mathcal{R}\sigma_{x}\mathcal{W}_{B}W_{B}U^{\dagger}\boldsymbol{\beta}(k;2)\nonumber \\
 & =\mathcal{M}(k)\boldsymbol{\beta}(k;2).\label{eq:bc_eigenbasis-1}
\end{align}
and the bulk equations,\begin{subequations}\label{eq:tB_diagonal_chain-1}
\begin{align}
\alpha_{\sigma}(k;n+1)-\xi_{A}^{\sigma}\alpha_{\sigma}(k;n) & =-\left(\frac{\epsilon}{t}\right)\beta_{\sigma}(k;n+1),\label{eq:tB_diagonal_chain_A-1}\\
\beta_{\sigma}(k;n)-(\xi_{A}^{\sigma})^{*}\beta(k;n+1) & =-\left(\frac{\epsilon}{t}\right)\alpha_{\sigma}(k;n).\label{eq:tB_diagonal_chain_B-1}
\end{align}
\end{subequations} The matrix $\mathcal{M}(k)$ can be calculated
explicitly, since all the matrices intervening in its definition were
given above, but its long expression is not particularly enlightening.


\section{The low-energy edge states}

\label{sec:LowEnEdgeStates}

We now sketch the calculation of the low energy edge states for the
problem set by Eqs.(\ref{eq:tB_matrix}) in a semi-infinite ribbon.
For solutions that decay away from the edge, \begin{subequations}\label{eq:edge4_solution-1}
\begin{align}
\alpha_{\sigma}(k;n) & =e^{iq_{\sigma}(n-2)}\alpha_{\sigma}(k;q_{\sigma}),\label{eq:edge4_solution_alpha-1}\\
\beta_{\sigma}(k;n) & =e^{iq_{\sigma}(n-2)}\beta_{\sigma}(k;q_{\sigma}),\label{eq:edge4_solution_beta-1}
\end{align}
\end{subequations} the equations for the amplitudes in the bulk become
\begin{subequations}\label{eq:bulk_eqs_ref}

\begin{align}
(1-e^{-iq_{\sigma}}\xi_{A}^{\sigma})\alpha_{\sigma}(k;q^{\sigma}) & =-\left(\frac{\epsilon}{t}\right)\beta_{\sigma}(k;q_{\sigma}),\label{eq:bulk_eqs_ref-a}\\
(1-e^{iq_{\sigma}}(\xi_{A}^{\sigma})^{*})\beta_{\sigma}(k;q^{\sigma}) & =-\left(\frac{\epsilon}{t}\right)\alpha_{\sigma}(k;q_{\sigma}).\label{eq:bulk_eqs_ref-b}
\end{align}
\end{subequations}The energy must be given by 
\begin{align}
\left(\frac{\epsilon}{t}\right)^{2} & =\left(1-e^{-iq_{\sigma}}\xi_{A}^{\sigma}\right)\left(1-e^{+iq_{\sigma}}(\xi_{A}^{\sigma})^{*}\right).\label{eq:TB_EnergyDisp}
\end{align}
Expanding the RHS, and given the fact that the energy must be real,
we conclude that $\Im\left[e^{-i\Re q_{\sigma}}\xi_{A}^{\sigma}\right]=0$,
which is equivalent to $e^{-iq_{\sigma}}\xi_{A}^{\sigma}=\pm\left|\xi_{A}^{\sigma}\right|e^{\Im q^{\sigma}}$.
This allows us to rewrite Eq. (\ref{eq:TB_EnergyDisp}) as

\begin{align}
\left(\frac{\epsilon}{t}\right)^{2} & =1+\left|\xi_{A}^{\sigma}\right|^{2}\mp2\left|\xi_{A}^{\sigma}\right|\cosh\left(\Im q^{\sigma}\right).\label{eq:TB_EnergyDisp-2}
\end{align}
Low energy solutions, with $\left|\epsilon/t\right|\ll1$, correspond
to the choice of the minus sign in this expression. From this, we
can write the energy expression as
\begin{eqnarray}
\frac{\epsilon}{t} & = & -(1-\left|\xi_{A}^{\sigma}\right|e^{\Im q^{\sigma}})\frac{1}{s_{\sigma}}.\label{eq:TB_nonSqEn}
\end{eqnarray}

On the other hand, the energy can be eliminated from Eqs. (\ref{eq:bulk_eqs_ref})
to obtain,
\begin{equation}
\frac{1-e^{\Im q_{\sigma}}\left|\xi_{A}^{\sigma}\right|}{1-e^{-\Im q_{\sigma}}\left|\xi_{A}^{\sigma}\right|}=\left(\frac{\beta_{\sigma}(k,q_{\sigma})}{\alpha_{\sigma}(k,q_{\sigma})}\right)^{2}:=s_{\sigma}^{2}\label{eq:TB_Imq}
\end{equation}

This result shows that the values of $\Im q_{\sigma}$ are determined
if we fix the amplitude ratios, $s_{\sigma}$, \emph{i.e.,} if we
take as BCs for the two $\sigma=+,-$, chains 
\[
\beta_{\sigma}(k,q_{\sigma})=s_{\sigma}\alpha_{\sigma}(k,q_{\sigma}).
\]

To determine the value of the energy we use the following conditions:
(i) the values of $s_{+}$and $s_{-}$ are related by the BCs {[}Eq.
(\ref{eq:bc_eigenbasis-1}){]}, 
\begin{eqnarray}
\frac{1}{s_{+}} & = & \frac{\mathcal{M}_{++}-\det[\mathcal{M}]s_{-}}{1-\mathcal{M}_{--}s_{-}};\label{eq:ref_bc-1}
\end{eqnarray}
 (ii) their values must be such that the RHS of Eq. (\ref{eq:TB_Imq})
is independent of $\sigma.$ Hence, we determine $\Im q_{-}$and $\Im q_{+}$,
as a function of $s_{-}$ (using the value of $s_{+}$ given by Eq.
(\ref{eq:ref_bc-1}), calculate the energies from Eq. (\ref{eq:TB_nonSqEn})
for $\sigma=+,-$, and vary $s_{-}$ until the two energies match;
as long as $\left|e^{\Im q^{\sigma}}\right|<1$, this constitutes
the solution of our problem.

Note that the sign of the energy, is determined by the hopping amplitudes
trough Eq. (\ref{eq:ref_bc-1}). The BCs we used are only valid for
$\left|\epsilon/t\right|\ll1$. As a consequence, we can expect that
this analytical construction of edge states will only be valid near
the Dirac points ($ka=\pm\pi/3$), where this condition is fulfilled.


\section{Recurrence matrices with magnetic field}

\label{sec:MatricesAppndxB1}

When a perpendicular magnetic field is applied perpendicularly to
the ribbon, in the bulk, the matrices $\widetilde{W}_{A}(n)$ and
$\widetilde{W}_{B}(n)$ read \begin{subequations} \label{eq:WsMatricesB}
\begin{eqnarray}
\widetilde{W}_{A}(n) & = & -\left[\begin{array}{cc}
e^{i(n+1)\pi\frac{\phi_{6}}{\phi_{0}}} & e^{-i(n+1)\pi\frac{\phi_{6}}{\phi_{0}}}\\
e^{i\left(2ka-(n+1)\pi\frac{\phi_{6}}{\phi_{0}}\right)} & e^{i(n+1)\pi\frac{\phi_{6}}{\phi_{0}}}
\end{array}\right],\nonumber \\
\label{eq:tWa}\\
\widetilde{W}_{B}(n) & = & -\left[\begin{array}{cc}
e^{-i(n+1)\pi\frac{\phi_{6}}{\phi_{0}}} & e^{-i\left(2ka-(n+1)\pi\frac{\phi_{6}}{\phi_{0}}\right)}\\
e^{i(n+1)\pi\frac{\phi_{6}}{\phi_{0}}} & e^{-i(n+1)\pi\frac{\phi_{6}}{\phi_{0}}}
\end{array}\right],\nonumber \\
\label{eq:tWb}
\end{eqnarray}
 \end{subequations} where $\phi_{6}$ is the magnetic flux through
an undistorted hexagon.

Moreover, the matrices around the upper edge,~$\widetilde{\mathcal{W}}_{A}^{U}$,
$\widetilde{\mathcal{W}}_{B}^{U}$ and $\widetilde{\mathcal{R}}$,
are given by\begin{subequations} \label{eq:MatricesB} 
\begin{eqnarray}
\widetilde{\mathcal{W}}_{A}^{U} & = & -\left[\begin{array}{cc}
h'_{1}e^{i\pi\frac{\phi_{d6}^{U}}{\phi_{0}}} & h'_{1}e^{-i\pi\frac{\phi_{d6}^{U}}{\phi_{0}}}\\
h'_{3}e^{i(2ka-\pi\frac{\phi_{d6}^{U}}{\phi_{0}})} & h'_{3}e^{i\pi\frac{\phi_{d6}^{U}}{\phi_{0}}}
\end{array}\right],\label{eq:WaB}\\
\widetilde{\mathcal{W}}_{B}^{U} & = & -\frac{1}{v}\left[\begin{array}{cc}
h'_{1}e^{-i\pi\frac{\phi_{d6}^{U}}{\phi_{0}}} & h'_{3}e^{-i(2ka-\pi\frac{\phi_{d6}^{U}}{\phi_{0}})}\\
h'_{1}e^{i\pi\frac{\phi_{d6}^{U}}{\phi_{0}}} & h'_{3}e^{-i\pi\frac{\phi_{d6}^{U}}{\phi_{0}}}
\end{array}\right],\label{eq:WbB}\\
\widetilde{\mathcal{R}} & = & -\left[\begin{array}{cc}
\frac{h_{1}^{2}-h_{4}h_{2}e^{-i\theta}}{vh_{2}} & 0\\
0 & \frac{h_{1}^{2}-h_{4}h_{2}e^{i\theta}}{vh_{2}}
\end{array}\right],\label{eq:RRB}
\end{eqnarray}
\end{subequations} where $\theta=2ka-\pi\frac{\phi_{5}^{U}}{\phi_{0}}-\pi\frac{\phi_{7}^{U}}{\phi_{0}}$
and $\phi_{7}^{U}$, $\phi_{5}^{U}$ and $\phi_{d6}^{U}$ are the
fluxes of the magnetic field across the upper heptagons, pentagons
and distorted hexagons (see Fig. \ref{fig:PeierlsPhases}), while
$\phi_{0}$ is the flux quantum. The matrix associated with the boundary
at the upper edge, $\widetilde{\sigma_{x}}$, reads
\begin{eqnarray}
\widetilde{\sigma}_{x} & = & \left[\begin{array}{cc}
0 & e^{i\pi\frac{\phi_{7}^{U}}{\phi_{0}}}\\
e^{-i\pi\frac{\phi_{7}^{U}}{\phi_{0}}} & 0
\end{array}\right].\label{eq:sigB}
\end{eqnarray}

If we take the energy to be zero, and change to the proper basis,
the BC for the edge at $n=0$ becomes
\begin{eqnarray}
\mathbf{\boldsymbol{\alpha}}(k;2) & = & -U\widetilde{\mathcal{W}}_{A}^{U}\widetilde{\mathcal{R}}\widetilde{\sigma_{x}}\widetilde{\mathcal{W}}_{B}^{U}\widetilde{W}_{B}(2)U^{\dagger}\mathbf{\boldsymbol{\beta}}(k;2)\nonumber \\
 & = & \widetilde{\mathcal{M}}(k)\boldsymbol{\beta}(k;2).\label{eq:BC_MagF}
\end{eqnarray}

The proper basis of matrices $\widetilde{W}_{A}$ and $\widetilde{W}_{B}$
is $\{\boldsymbol{u}^{+},\boldsymbol{u}^{-}\}$ defined in Appendix
\ref{sec:appendix}. In the proper basis, the equations for the bulk
amplitudes, read\begin{subequations}\label{eq:tB_diagonal_chain_MagF}

\begin{eqnarray}
-\alpha_{\sigma}(k;n+1)+\widetilde{\xi_{A}^{\sigma}}(n)\alpha_{\sigma}(k;n) & = & -\left(\frac{\epsilon}{t}\right)\beta_{\sigma}(k;n),\nonumber \\
\\
-\widetilde{\xi_{B}^{\sigma}}(n)\beta_{\sigma}(k;n)+\beta_{\sigma}(k;n-1) & = & -\left(\frac{\epsilon}{t}\right)\alpha_{\sigma}(k;n),\nonumber \\
\end{eqnarray}
\end{subequations}where the $\widetilde{\xi_{A/B}^{\sigma}}$ are
defined in Eqs. (\ref{eq:xiGen}).

\bibliographystyle{apsrev}


\end{document}